\documentclass[twocolumn]{aastex63}
\newcommand{\kms}{\,km\,s$^{-1}$} 
\newcommand{\Av}{$A\rm _{V}$}

\submitjournal{APJ}

\shorttitle{Mapping dust attenuation in galaxies}
\shortauthors{Zhou et al.}

\begin{document}

\title{Mapping dust attenuation and the 2175 {\AA} bump at kpc scales in nearby galaxies}

\email{Contact e-mail: shuang.zhou@nottingham.ac.uk (SZ);\\cli2015@tsinghua.edu.cn (CL)}

\author[0000-0002-8999-6814]{Shuang Zhou}
\affiliation{Department of Astronomy, Tsinghua University, Beijing 100084, China}
\affiliation{School of Physics \& Astronomy, University of Nottingham, University Park, Nottingham, NG7 2RD, UK}

\author[0000-0002-8711-8970]{Cheng Li}
\affiliation{Department of Astronomy, Tsinghua University, Beijing 100084, China}

\author[0000-0002-0656-075X]{Niu Li}
\affiliation{Department of Astronomy, Tsinghua University, Beijing 100084, China}

\author[0000-0001-5356-2419]{Houjun Mo}
\affiliation{Department of Astronomy, University of Massachusetts Amherst, MA 01003, USA}

\author[0000-0003-1025-1711]{Renbin Yan}
\affiliation{Department of Physics, The Chinese University of Hong Kong, Sha Tin, NT, Hong Kong, China}

\author[0000-0002-3719-940X]{Michael Eracleous}
\affiliation{Department of Astronomy and Astrophysics, The Pennsylvania State University, 525 Davey Lab, University Park, PA 16802, USA}
\affiliation{Institute for Gravitation and the Cosmos, The Pennsylvania State University, University Park, PA 16803, USA}

\author[0000-0001-8440-3613]{Mallory Molina}
\affiliation{Department of Physics \& Astronomy, University of Utah, James Fletcher Building, 115 1400 E, Salt Lake City, UT 84112, USA}

\author[0000-0001-6842-2371]{Caryl Gronwall}
\affiliation{Department of Astronomy and Astrophysics, The Pennsylvania State University, 525 Davey Lab, University Park, PA 16802, USA}
\affiliation{Institute for Gravitation and the Cosmos, The Pennsylvania State University, University Park, PA 16803, USA}

\author[0000-0003-1469-8246]{Nikhil Ajgaonkar}
\affiliation{Department of Physics and Astronomy, University of Kentucky, 505 Rose St., Lexington, KY 40506-0057, USA}

\author[0009-0001-5437-410X]{Zhuo Cheng}
\affiliation{Department of Astronomy, Tsinghua University, Beijing 100084, China}

\author[0009-0008-4962-665X]{Ruonan Guo}
\affiliation{Department of Astronomy, Tsinghua University, Beijing 100084, China}

\begin{abstract}
We develop a novel approach to measure dust attenuation properties of galaxies, including the dust opacity, shape of the attenuation curve and the strength of the 2175{\AA} absorption feature. From an observed spectrum, the method uses a model-independent approach to derive a relative attenuation curve, with absolute amplitude calibrated using NIR photometry. The dust-corrected spectrum is
fitted with stellar population models to derive the dust-free model spectrum, which is compared with the observed SED/spectrum from NUV to NIR to determine dust attenuation properties. We apply this method to investigate dust attenuation on kpc scales, using a sample of 134 galaxies with integral field spectroscopy from MaNGA, NIR imaging from 2MASS, and NUV imaging from Swift/UVOT. We find the attenuation curve slope and the 2175{\AA} bump in both optical and NUV span a wide range at kpc scales. The slope is shallower at higher optical opacity, regardless of the specific star formation rate (sSFR), minor-to-major axis ratio (b/a) of galaxies and the location of spaxels within individual galaxies. The 2175{\AA} bump presents a strong negative correlation with the sSFR, while the correlations with the optical opacity, b/a and the location within individual galaxies are all weak. All these trends appear to be independent of the stellar mass of galaxies. Our results support the scenario that the variation of the 2175{\AA} bump is driven predominantly by processes related to star formation, such as the destruction of small dust grains by UV radiation in star-forming regions.   
\end{abstract}


\section{Introduction}

Interstellar dust is an important component of interstellar media, 
although it accounts for only a tiny fraction of the baryon mass 
\citep[e.g.][]{Remy-Ruyer-2014, Driver-2018}.  The light emitted by 
stars may be absorbed or scattered by dust particles, causing 
dust extinction and attenuation.  Dust extinction represents the loss
of starlight, which can be measured directly using resolved sources, 
such as stars with known intrinsic spectra. Limited by 
telescope resolution, such measurements can be achieved only for the Milky
Way (MW; \citealt{Fitzpatrick1999}) and a few nearby galaxies, such as
the Large Magellanic Cloud (LMC, \citealt{Gordon2003}), the Small
Magellanic Cloud (SMC, \citealt{Gordon1998}) and M31
\citep[e.g.][]{Clayton2015}. For distant galaxies, however, the
redistribution of their spectral energy distribution (SED) due to
dust attenuation is much more complex. The shape of the dust
attenuation curve (i.e., attenuation as a function of wavelength)
depends on both the intrinsic properties of dust particles and the
distributions of the dust and stellar contents along the line of sight (see
\citealt{Calzetti2001review} for a review). To get the intrinsic
SED of a galaxy, one must carefully model the effects of dust and
make proper correction for them. So far, the attenuation
curve that should be used in the correction has not been identified
and remains elusive in observational studies of galaxy spectra.

\citet{Calzetti1994,Calzetti2000} analysed a sample of local star
burst galaxies and derived an average attenuation curve, often
referred to as the Calzetti attenuation curve. \cite{Charlot2000}
proposed a two-component dust model characterized by an attenuation
curve that is proportional to $\lambda^{-0.7}$ but with different
optical depth for star-birth clouds and diffuse interstellar media
(ISM), in order to reproduce both the infrared/ultraviolet luminosity
ratio and the ultraviolet (UV) spectral slope. 
Both the Calzetti and \citet{Charlot2000} curves
have similar shapes in the optical band and both have been widely adopted
in spectral and SED analysis of galaxies. However, there are 
a number of unsolved problems, especially when near ultra-violet (NUV) 
observations are taken into account in the analysis. For example, 
observations of the  Milky Way (MW) show a strong absorption feature at the rest-frame wavelength 
around 2175{\AA} \citep[e.g.][]{Cardelli1989,Fitzpatrick1999}, which is often referred
to as the 2175{\AA} bump. Similar absorption features, although with
varying strength, are also seen in several local galaxies, such as LMC
\citep{Gordon2003} and M31 \citep[e.g.][]{Clayton2015}. However,
this feature appears to be negligible in the sample of starburst galaxies
used by \citet{Calzetti1994}.

Large multi-band surveys conducted in the past two decades have
allowed statistical analyses of the 2175{\AA} attenuation feature using
large samples of galaxies (see the review by \citealt{Salim2020}).  At
low redshifts, most of the studies have relied on photometry due to
the lack of rest-frame UV spectra.  For instance, \citet{Conroy2010dust}
analyzed a sample of disk-dominated star-forming galaxies with
photometry available in both UV from the Galaxy Evolution Explorer
\citep[{\it GALEX};][]{GALEX} and optical from the Sloan Digital Sky Survey
\citep[SDSS;][]{York2000}. They found that the Calzetti curve provides poor 
fits to ultraviolet colors for moderately and highly inclined galaxies, 
and they speculated that the existence of the
2175{\AA} bump is responsible for the observed trends in their
galaxies. Using the near-infrared photometry from the UKIRT Infrared Deep 
Sky Survey-Large Area Survey \citep[UKIDSS-LAS;][]{UKIDSS} and emission 
lines from the SDSS spectroscopic survey, 
in combination with {\it GALEX} and SDSS photometric data,
\citet{Wild2011} found some evidence for the existence of the UV absorption
feature in low-redshift star-forming galaxies. In contrast,
\cite{Battisti2016} found no evidence for the 2175{\AA} feature
from a sample of $\sim10,000$ local star forming galaxies from 
SDSS and {\it GALEX}. More recently, \cite{Salim2018} derived dust attenuation curves 
by applying the SED fitting code {\tt CIGALE} to $\sim$230,000 star-forming 
galaxies with photometry from {\it GALEX}, SDSS, and  the Wide-field Infrared Survey
Explorer \citep[WISE;][]{WISE}. They found that the dust curves of individual 
galaxies have large variances, but that the average curve is 
similar to that given by \citet{Conroy2010dust} and that the average UV bump 
strength is about one third of the MW value.

At higher redshifts where the rest-frame NUV falls in the optical, 
analyses of the UV absorption feature can be carried using photometric 
and spectroscopic data in the optical band.  
For instance, a series of studies of  massive star forming galaxies at 
$1 < z < 2.5$ indicated 
the existence of a UV bump of moderate strength
\citep{Noll2005,Noll2007,Noll2009dust}. A later study by
\cite{Buat2011,Buat2012} also suggested that a UV bump with a strength
$\sim35\%$ of the MW value is needed to model the SED of star forming galaxies 
at $1<z < 2.2$. \cite{Kriek2013} detected the presence of a UV bump using 
a sample of galaxies at $0.5 < z < 2.0$, and found a correlation between the bump 
strength and the slope of the attenuation curve. \cite{Scoville2015} reported a UV 
bump feature in galaxies at even higher redshifts ($z\sim2-6.5$). More recent studies based on
different surveys \citep{Reddy2015,Leja2017,Shivaei2020,Barisic-2020,Battisti2020,Kashino2021}
generally yielded results that are consistent with the existence of the bump, 
with the exception of \citet{Zeimann2015}, but the bump strengths obtained  
from different samples seemed to vary significantly.

Despite of a long history of investigation
\citep[e.g.][]{Stecher1965,Savage1975,Joblin1992,Beegle1997}, the dust
species that dominates the bump at 2175{\AA} is still under debate
\citep[][]{Bradley2005,Papoular2009,Steglich2010}.  The fact that the UV bump 
strength in observed dust attenuation curves vary from galaxy to galaxy indicates 
either that those galaxies are intrinsically different in dust particle 
properties, or that the differences are caused by radiative transfer effects 
in a complex star-dust distribution in galaxies.   
Analyses of the evolution of the distribution of
dust grains \citep[e.g.][]{Asano2013,Asano2014,Hirashita2015} indicated 
that the 2175{\AA} bump strength may be linked to the relative
abundance of small dust grains in galaxies,  while the lack of a
2175{\AA} bump in starburst galaxies may be explained by the
destruction of such grains \citep{Fischera2011}. 
On the other hand, models incorporating radiative transfer effects 
and stellar-dust geometry \citep[e.g.][]{Gordon1997,Witt2000,Seon2016,Narayanan2018} 
suggest that the dust attenuation curve is shallower as the 2175\AA~ bump 
weakens (thus more similar to the Calzetti curve), and the dependence 
becomes stronger as the optical depth increases. 
Spatially resolved observations down to scales of star-forming regions 
and  covering both optical and UV bands are needed to better understand
the driving processes for the variation of the dust attenuation curve in
galaxies.

In this paper we investigate the dust attenuation for a sample of
134 galaxies in the local Universe, using optical integral-field
spectroscopy from the Mapping Nearby Galaxies at Apache Point
Observatory \citep[MaNGA;][]{Bundy2015} survey, NUV photometry
obtained by {\it Swift}/UVOT \citep{Roming2005}, and NIR photometry from the
Two-Micron All Sky Survey \citep[2MASS;][]{Skrutskie2006}. Different
from  previous studies based either on the global photometry
of galaxies or on single-fiber spectroscopy limited to the central part
of galaxies, MaNGA provides spatially-resolved spectroscopy to 
investigate dust attenuation down to kpc scales in individual galaxies, 
thus effectively reducing the influence of dust distribution in galaxies. By
combining MaNGA, {\it Swift}/UVOT and 2MASS data, not only can we resolve  
dust attenuation spatially, but also probe it over a wavelength 
range from NUV to NIR. In addition, our method takes
advantage of a newly developed technique
\citep{Li2020} which estimates the relative attenuation curve from
spectral fitting without the need of assuming a functional
form for the attenuation curve. Here we extend the
technique by calibrating the absolute amplitude of
the attenuation curve with NIR photometry, and 
by incorporating the UV photometry from {\it Swift}/UVOT
to study the bump around 2175{\AA} in the dust curve.  
This method provide measurements not only for the UV bump strength, 
but also for the slope of the attenuation curve in both the optical 
and UV bands. The relatively large sample also allows us 
to statistically study the correlation of dust properties and their dependence 
on star formation rate.

The paper is organized as follows. We describe our data in
\S\ref{sec:data}, and present the method used to estimate dust
attenuation curves in \S\ref{sec:method}.  We present our results
in \S\ref{sec:results}, and our discussion
in \S\ref{sec:discussion}. Finally we summarize in 
\S\ref{sec:summary}. A standard $\Lambda$CDM cosmology with
$\Omega_{\Lambda}=0.7$, $\Omega_{\rm M}=0.3$ and $H_0$=70\kms
Mpc$^{-1}$ is assumed throughout the paper. 

\section{Data}
\label{sec:data}

\subsection{MaNGA}
\label{sec:manga}

As one of the major experiments of the fourth-generation SDSS project
\citep[SDSS-IV;][]{Blanton2017}, MaNGA has successfully obtained
integral field spectroscopy (IFS) for 10,010 nearby galaxies during
the period from July 2014 through August 2020 \citep{Bundy2015}. The
IFS data are obtained with integral field units (IFUs) with various
sizes. Each IFU is a hexagonal-formatted fiber bundle made from
2$^{\prime\prime}$-core-diameter fibers with 0.5$^{\prime\prime}$ gaps
between adjacent fiber cores. When combined with the seeing, they can provide IFS datacubes with an
effective spatial resolution that can be described by a Gaussian with
a full width at half maximum (FWHM) of $\sim2.5^{\prime\prime}$
\citep{Drory2015,Law2015}.  The IFU fibers are fed to the two
dual-channel BOSS spectrographs on the Sloan 2.5-metre telescope
\citep{Gunn2006,smee2013} to obtain MaNGA spectra in
a wavelength range from 3622{\AA} to 10354{\AA} with a spectral
resolution of $R\sim2000$. With a typical exposure time of about 
three hours, the observational data reach a $r$-band signal-to-noise
(S/N) of 4-8 per {\AA} per $2^{\prime\prime}$ fiber at 1-2 effective
radii ($R_e$) of galaxies, 

MaNGA galaxy targets are selected from an updated version of the NASA
Sloan Atlas catalogue \citep[NSA\footnote{\label{foot:nsa}\url{
 http://www.nsatlas.org/}};][]{Blanton2011}.  Three samples are
selected: Primary, Secondary, and Color-Enhanced samples. As the main
samples of the survey the Primary and Secondary samples have a flat
distribution in the $K$-corrected $i$-band absolute magnitude ($M_i$)
with the assigned IFUs covering out to 1.5 and 2.5$R_e$ of the
galaxies, respectively. The Color-Enhanced sample additionally selects
galaxies on the $NUV-r$ versus $M_i$ diagram that are not well sampled
by the Primary and Secondary samples. Overall, these samples cover a
wide range of stellar mass ($10^9$M$_{\odot}\lesssim~M_\ast\lesssim6\times10^{11}$M$_{\odot}$)
and a redshift range of $0.01<z<0.15$, with a median redshift
$z\sim0.03$ \citep{wake2017}.  MaNGA raw data are reduced with  the
Data Reduction Pipeline \citep[DRP;][]{Law2016,Law2021} to produce a
datacube for each galaxy with a spaxel size of
0.5$^{\prime\prime}\times$0.5$^{\prime\prime}$.  The absolute flux
calibration of the MaNGA spectra is better than 5\% for more than 80\%
of the wavelength range. The flux calibration, MaNGA survey strategy
and data quality tests are described in detail by \cite{Yanb2016} and
\cite{Yana2016}. In addition, the MaNGA Data Analysis Pipeline
\citep[DAP,][]{Westfall2019,Belfiore2019} provides  measurements of
stellar kinematics, emission lines and spectral indices, obtained by
performing full spectral fitting to the DRP datacubes. All the MaNGA
data, including DRP and DAP products of the 10,010 galaxies, are
released as part of the final data release of SDSS-IV
\citep[DR17\footnote{\url{https://www.sdss.org/dr17/manga/}};][]{SDSS_DR17}.

\subsection{SwiM}
\label{sec:swim}

We use the {\it Swift}/UVOT+MaNGA (SwiM) value-added
Catalog \citep{Molina2020a} which includes 150 galaxies with public
MaNGA data released with the SDSS Data Release 15 \citep{Aguado2019}
and NUV imaging data from the {\it Swift}/UVOT NUV archive as of April 26,
2018. The Ultraviolet Optical Telescope (UVOT; \citealt{Roming2005})
is one of the three instruments on the {\it Swift} Gamma-ray Observatory.
It has a $17^\prime\times17^\prime$ field-of-view (FOV) and
operates in the wavelength range of 1600–8000{\AA}. Imaging data are
taken in three NUV bands: {\tt uvw2}, {\tt uvm2} and {\tt uvw1}
centered at 1928, 2246 and  2600{\AA} respectively, with a point
spread function (PSF) of around 2.5$^{\prime\prime}$ that is similar
to the MaNGA spatial resolution. Note that the  {\tt uvm2}  filter is centered 
near the 2175{\AA} bump feature as seen in the Milky Way dust curve, 
making it suitable for the investigation of such a feature (see \autoref{method_bump}).  
All the 150 galaxies in the SwiM
catalog have data from MaNGA and in the {\tt uvw1} and {\tt uvw2} bands,
and 134 of them  also have data in the {\tt uvm2}
band.  As detailed by \citet{Molina2020a}, all the {\it Swift}/UVOT images
are reprocessed carefully. For each galaxy the {\it Swift}/UVOT images
in {\tt uvw1} and {\tt uvm2} and the MaNGA datacube are convolved and
resampled to match the spatial resolution and saptial sampling in  the
{\tt uvw2} band that has the coarsest PSF (2.92$^{\prime\prime}$
FWHM) and a pixel size of $1^{\prime\prime}$. \citet{Molina2020a}
found that the distribution of SwiM galaxies in color, effective radius 
and axial ratio is similar to that of the MaNGA sample, but
is slightly biased for the population of low stellar mass and low
star formation rate. The reader is referred to \citet{Molina2020a}
for more details about the SwiM catalog.  
A subset of 29 galaxies in the SwiM catalog has been 
used to study the UV stellar attenuation and optical nebular attenuation 
\citep{Molina2020b} as well as the relationship between the infrared excess 
(IRX) and the UV spectral index \cite{Duffy2023}
 in kpc-sized star forming regions.
Here we consider all the 134 galaxies that have data in all the
three NUV bands, thus including both star-forming and quiescent galaxy
populations. 

\subsection{2MASS}

We also use near-infrared images from the
2MASS \citep{Skrutskie2006}.  We retrieved Ks-band atlas images of the
134 SwiM galaxies from the 2MASS data Archive. The 2MASS images have
the same sampling (1$^{\prime\prime}$ per pixel) and a very similar
PSF ($\sim$2.5-3.5$^{\prime\prime}$ FWHM) in comparison to the SwiM
catalog (see above). We tried to convolve images from the three UVOT bands and 
the 2MASS Ks band so that they all match the lowest resolution among them. 
Since the spatial resolutions are not very different, the flux differences 
in pixels before and after the convolution are typically less than $1\%$, 
much smaller than observational uncertainties. 
We thus decide to simply re-sample the 2MASS images to match the spatial 
sampling of the SwiM images, without performing additional resolution matching. 

\section{Methodology}
\label{sec:method}

We estimate dust attenuation properties for each spatial pixel
(spaxel) of our galaxies.  These include (1) an absolute dust
attenuation curve in the optical, $A_{\rm opt}(\lambda)$, (2) the
slope of the attenuation curve in NUV as defined by the ratio of the
attenuation between the {\tt uvw2} and {\tt uvw1} bands, $A_{\tt
  w2}/A_{\tt w1}$, and 
(3) the 2175{\AA} bump, characterized by 
the extra attenuation at 2175{\AA} due to the bump ($A_{\text{bump}}$) 
and the strength of the bump ($B$).
We obtain these measurements in  two successive steps. First, we
obtain $A_{\rm opt}(\lambda)$ as well as a dust-free model spectrum
covering the full wavelength range from NUV to NIR, using the MaNGA
spectrum and the Ks band photometry in the spaxel. Next, the dust-free
model spectrum in the NUV is combined with the {\it Swift}/UVOT photometry
in {\tt uvw2}, {\tt uvm2} and {\tt uvw1} to estimate $A_{\tt
  w2}/A_{\tt w1}$, $A_{\text{bump}}$ and $B$.  In the rest of this section, 
we describe the two steps in detail and test our method with a set of mock
spectra. 

\subsection{Deriving the optical attenuation curve and the dust-free model spectrum}
\label{sec:attenuation_curve}

\begin{figure*}
    \includegraphics[width=0.98\textwidth]{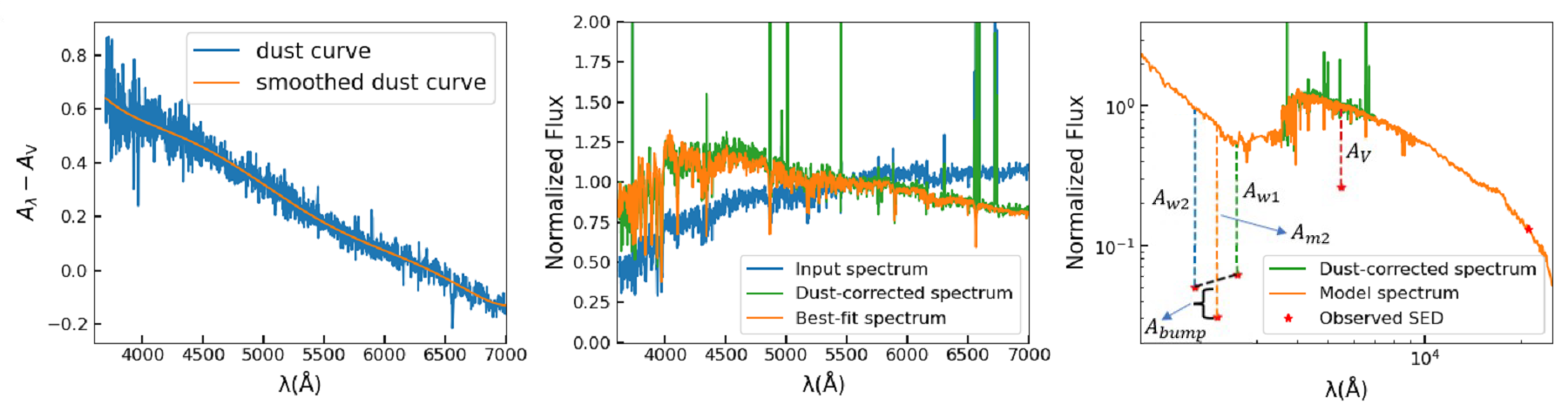}
    \caption{Illustration of the method used in this work. The blue line in the middle panel is one spectrum stacked within 1Re of our sample galaxy. In the plot we normalized the spectrum at 5500\AA. We use the method of \citealt{Li2020} to fit the spectrum and correct for the dust attenuation in optical range. The method of  \citealt{Li2020} will give a relative attenuation curve, which is shown in the left panel. After correcting for the dust attenuation, we fit the corrected spectrum with the spectrum fitting code {\tt BIGS}. As shown in the right plot, the best-fit spectrum provided by {\tt BIGS} is extended to both NUV and NIR (orange line). This dust-free spectrum is convolved with filter response functions to predict dust-free SED, which is compared with the observed SED shown by red stars in the right plot. By assuming that the Ks band is not affected by dust attenuation, the observed SED is also normalized so that the flux at Ks band is the same as the dust-free spectrum. The difference between the dust-free SED and observed SED is our attenuation estimation, which is indicated as dash lines in the right panel.}
    
    \label{fig:method_demo}
\end{figure*}

For a given spaxel of each galaxy, we start by  applying the technique
of \citet{Li2020} to the MaNGA spectrum to estimate a relative
attenuation curve in the optical range, $A_{\rm opt}(\lambda)-A_{\rm
  V}$, where $A_{\rm V}$ is the dust attenuation in the V-band.  The
attenuation curve is then used to correct the effect of relative
attenuation in the observed spectrum, resulting in a dust-free
spectrum but with an arbitrary flux unit. Next, a spectral
fitting code, Bayesian Inference of Galaxy Spectra ({\tt BIGS}; \citealt{Zhou2019}), 
is applied to the dust-free spectrum to obtain a best-fit model spectrum covering the
full wavelength range from NUV to NIR. Finally, this model spectrum is
flux calibrated  using the Ks band photometry which is assumed to
be unaffected by the dust.

The detail of our method to estimate $A_{\rm opt}(\lambda)-A_{\rm V}$
can be found in \citet{Li2020}. 
In short, for a given spectrum, this method first decomposes it into two 
	components:  one representing small-scale features of the spectrum ($S$) 
	and one describing the large-scale spectral shape ($L$). 
	The large-scale component $L$ is obtained by applying a 
	moving average filter:
\begin{equation}\label{eq:large}
  F_L(\lambda )= \frac{1}{\Delta \lambda}
  \int_{\lambda-\Delta \lambda/2}^{\lambda+\Delta \lambda/2}
  F(\lambda')\,d\lambda',
\end{equation}
where F and F$_L$ are the original spectrum and the $L$ component respectively, 
	and $\Delta \lambda$ specifies the size of the wavelength window of the filter. 
	The spectrum of the small-scale component is then given by
\begin{equation}\label{eq:small}
  F_S(\lambda)=F(\lambda)-F_L(\lambda).
\end{equation}
This decomposition is performed for both the observed spectra and the 
spectra of all the simple-stellar-population (SSP) models of \citet[][hereafter BC03]{BC03}. 
As shown in \cite{Li2020} (see \S2.1 and Eqns. 3-8 in that paper), the two
components are attenuated by dust in the same way and so their ratio
$S/L$ is expected to be free of dust attenuation, as long as
dust attenuation curves are similar for different stellar populations 
or the optical depths are smaller than unity.  Therefore, the intrinsic 
dust-free spectrum of the stellar populations in a given spaxel
can be derived by fitting the $S/L$ of the observed spectrum to the 
$S/L$ of the SSP models. 
The relative attenuation curve $A_{\rm opt}(\lambda)-A_{\rm
  V}$ is then obtained by comparing the observed spectrum with the
best-fit model spectrum.  One important advantage of this method is
that $A_{\rm opt}(\lambda)-A_{\rm V}$ can be obtained directly for 
the entire wavelength range of the observed spectrum without the need 
of assuming a functional form for the curve.  Extensive tests on mock
spectra in \cite{Li2020} showed that the method can accurately recover
the input attenuation curves. The method produces unbiased estimates of 
the average $E(B-V)$ for spectra of various S/N and $E(B-V)$ values. 
The standard deviation of $E(B-V)$ obtained from individual spectra
is relatively small, and decreases from $\sim0.1\,{\rm mag}$ for 
spectra of S/N$=5$ down to $<0.03$ mag for S/N larger than 
20 (see Fig.7 of \citealt{Li2020}). We note that an uncertainty of 
0.03 mag in $E(B-V)$ induces an uncertainty of $\sim$0.1 mag in $A_{\rm V}$ 
for the Calzetti attenuation curve. In what follows we use spaxels with 
S/N$>20$ in MaNGA.  

\autoref{fig:method_demo} (the leftmost panel) shows the relative
attenuation curve obtained this way for an example spaxel in our
sample.  As can be seen, the curve inherits all the noise features
from the observed spectrum\footnote{Note that emission lines are
  masked out from the observed spectrum so they are not present in the
  attenuation curve; see \citealt{Li2020} for details.}. We fit the
curve with a polynomial to obtain a smooth curve (plotted as the
yellow line in the figure), which is then used to correct the dust
attenuation effect in the observed spectrum. The middle panel of the
same figure displays both the observed and the dust-corrected
spectra, as well as the best-fit model spectrum obtained by 
applying  {\tt BIGS} to the dust-corrected one. 

In fitting the dust-corrected spectrum, we follow our previous studies
\citep{Zhou2020, Zhou2021} and use a $\Gamma$+B model to characterize
the star formation history (SFH). In this case, a $\Gamma$ function
is adopted to describe the long-term history of star formation rate
over cosmic time:
\begin{equation}
    \Psi(t)=\frac{1}{\tau\gamma(\alpha,t_0/\tau)} 
\left({t_0-t\over \tau}\right)^{\alpha-1}
    e^{-(t_0-t)/\tau}\,,
\label{gamma-sfh}
\end{equation}
where $\alpha$ and $\tau$ are free parameters determining the overall
shape of the SFH, $t_0$ is the present-day time (i.e. the age of the
universe, assumed to be 14Gyr), $t_0-t$ is thus the look-back time,
and  $\gamma(\alpha,t_0/\tau)\equiv \int_0^{t_0/\tau}
x^{\alpha-1}e^{-x}\,dx$ normalizes the SFH over the age of the
universe.  An additional burst component characterised by a SSP with
the burst time being a free parameter is added to account for possible
bursty events in the history.  Having been recently used to
explore the SFHs of low mass galaxies \citep{Zhou2020} and massive red
spirals \citep{Zhou2021} in MaNGA, this model has shown substantial
robustness and flexibility.  The SFH model is combined with  the BC03 
SSP model spectra assuming a stellar initial mass function (IMF)
of \citet{Chabrier2003} to generate composite model spectra with
various model parameters (coefficients of SSPs, $\alpha$, $\tau$,
etc.). Each model spectrum is then convolved with the stellar velocity
dispersion derived above from the procedure of deriving the relative
attenuation curve to account for stellar kinematics and instrumental
broadening. For a given set of model parameters ($\theta$), the model
spectrum is then compared with the dust-corrected spectrum to calculate
a $\chi^2$-like likelihood:
\begin{equation}
\label{likelyhood}
\ln {L(\theta)}\propto-\frac{1}{2}\sum_{i,j=1}^N\left(f_{\theta,i}-f_{D,i}\right)\left({\cal
M}^{-1}\right)_{ij}\left(f_{\theta,j}-f_{D,j}\right)\,
\end{equation}
where $f_{\theta, i}$ is the predicted flux at the $i$-th wavelength
pixel given the parameter set $\theta$, $f_{D, i}$ is the  flux of the
dust-corrected spectrum at the same wavelength, and $N$ is the total
number of wavelength pixels. The covariance matrix of the data, ${\cal
  M}_{ij}\equiv \langle \delta f_{D, i}\delta f_{D, j} \rangle$, is
assumed to be diagonal and specified by the error spectrum.  With a
flat prior and the above likelihood function,  {\tt BIGS} utilizes the
{\tt MULTINEST} sampler \citep{Feroz2009,Feroz2013} through a
\textsc{Python} interface \citep{Buchner2014} to sample the posterior
distributions of all the model parameters, which are stored and used
for subsequent analysis.

The best-fit model spectrum obtained for the example spaxel from the 
posterior distributions of the model parameters is plotted in yellow in 
the middle panel of \autoref{fig:method_demo}. In the rightmost panel
of the same figure, the observed and the best-fit model spectra are 
plotted again, but over a wider range of wavelength. As can be seen, the
model spectrum covers not only the optical range, but also extends to both 
NUV and NIR. Note, however, that the technique of
\citet{Li2020} yields dust attenuation curves that are relative. 
Thus, both the dust-corrected spectrum and the best-fit model spectrum 
are also relative, with arbitrary amplitudes to be determined.
Assuming that dust attenuation is negligible in NIR, 
we can obtain the absolute normalization of the model spectrum by matching 
the Ks-band magnitude of the spaxel in the 2MASS image with the model 
magnitude in the same band. This normalization scheme is used in 
\autoref{fig:method_demo} and in our following analysis.

\subsection{Characterizing the dust attenuation and the 2175{\AA} bump}
\label{method_bump}
Given the flux-calibrated dust-free model spectrum derived above, one
can obtain the unattenuated magnitude  for a specific  band by
convolving the model spectrum with the corresponding filter response
function, and thus the absolute attenuation in the band by comparing
the observed magnitude with the unattenudated magnitude.  Here, we
consider five bands: $B$ and $V$ in the optical and the three
{\it Swift}/UVOT bands ({\tt uvw2}, {\tt uvm2} and {\tt uvw1}).
Following common practice and for each spaxel with spectral $S/N>20$,
we have estimated the following quantities to characterize the dust
attenuation in both optical and NUV:
\begin{itemize}
\item $A_{\rm B}$, $A_{\rm V}$ --- stellar dust attenuation in $B$ and $V$ band.
\item $E(B-V)\equiv A_{\rm B}-A_{\rm V}$ --- color excess in the optical.
\item $R_{\rm V}\equiv A_{\rm V}/E(B-V)$ --- total-to-selective
  attenuation ratio in $V$ band. This parameter has been commonly used
  to characterize the slope of attenuation curves. For example, 
  $R_{\rm V}=4.05$ for a standard Calzetti curve, and
  $R_{\rm V}=3.1$ for a Milky Way-like curve \citep[e.g.][]{Cardelli1989}.
  \item $A_{\rm B}/A_{\rm V}$ --- optical slope of the attenuation curve
    defined following \cite{Salim2020}. We have $A_{\rm B}/A_{\rm V}=1.25$
    and 1.32 for the Calzetti and Milky Way curves, respectively.
\item $A_{\tt w2}$, $A_{\tt m2}$, $A_{\tt w1}$ --- stellar dust
  attenuation in NUV bands. 
\item $A_{\tt w2}/A_{\tt w1}$--- NUV slope of the attenuation curve,
  which is similar to, but not exactly the same as the definition in
  \cite{Salim2020} where the UV slope is defined as the ratio of
  extinctions at 1000{\AA} and 3000{\AA}. With our definition, we 
  have $A_{\tt w2}/A_{\tt w1}=1.19$  for the Calzetti attenuation curve and 
  $A_{\tt w2}/A_{\tt w1}=1.24$ for the Milky Way curve.
\end{itemize}
In the rightmost panel of \autoref{fig:method_demo}, the observed SED of 
the example spaxel is plotted as red stars, and the stellar attenuation parameters,  
$A_{\rm V}$, $A_{\tt w2}$, $A_{\tt m2}$ and $A_{\tt w1}$, are indicated by the
vertical dashed lines. 

Next, we estimate two more parameters to characterize
the 2175{\AA} bump. The first parameter is $A_{\rm bump}$, defined as 
\begin{equation}
  A_{\rm bump} \equiv A(2175{\mbox{\AA}}) - A^\prime(2175{\mbox{\AA}}),
\end{equation}
where $A(2175{\mbox{\AA}})$ is the total attenuation at 2175{\AA} and
$A^\prime(2175{\mbox{\AA}})$ is the attenuation at the same wavelength 
in the absence of a 2175{\AA} bump. Thus, $A_{\rm bump}$ is a measure of 
the {\em extra} attenuation around 2175{\AA} relative to that without 
a 2175{\AA} bump. Considering that the central wavelength
of the {\tt uvm2} band (2246{\AA}) is close to 2175{\AA}, we simply
use $A_{\rm m2}$ derived above to approximate $A(2175{\mbox{\AA}})$ in the
definition. The equation can thus be rewritten as
\begin{eqnarray}
  A_{\rm bump} & = & A_{\rm m2} - A^\prime_{\rm m2} \nonumber \\
  & = & \rm (uvm2-uvm2_0) - (uvm2^\prime - uvm2_0) \nonumber \\
  & = & \rm uvm2 - uvm2^\prime,
\end{eqnarray}
where all the quantities are defined in the {\tt uvm2} band, 
and a prime denotes the absence of a 2175{\AA} bump. The observed
magnitude ${\rm uvm2}$ and  the unattenuated magnitude ${\rm uvm2_0}$
can be readily measured from the observed and the dust-free model
spectrum. However, the expected magnitude in the absence of the 
2175{\AA} bump,  ${\rm uvm2^\prime}$, cannot be obtained in a straightforward
way.  We follow the generally-adopted assumption that, in the absence
of the 2175{\AA} bump, the UV stellar continuum can be well described by
a power-law of
$F(\lambda)\propto\lambda^{-\beta}$\citep[e.g.][]{Calzetti1994,
  Battisti2016}. With this assumption, we can use fluxes 
  in ${\rm uvw1}$ and ${\rm uvw2}$, both located far away from 2175{\AA}, 
  to determine a power-law continuum in NUV, which is then used to  
  estimate ${\rm uvm2^\prime}$. The power-law continuum and $A_{\rm bump}$ 
  derived this way are indicated in the right panel of \autoref{fig:method_demo}
  for the example spaxel. 

The second parameter is the bump strength, $B$, defined as follows 
	following common practice:
\begin{equation}
B \equiv A_{\rm bump}/A^\prime_{\rm m2}=A_{\rm bump}/(A_{\rm m2}-A_{\rm bump}).
\end{equation}
So defined, $B=0.19$ for a Milky Way-type dust curve, and 
$B=0$ for a Calzetti curve (no bump).  As the bump strength parameter is
defined relative to $A^\prime_{\rm m2}$, the uncertainty becomes large when $A^\prime_{\rm m2}$ is close to zero. Previous investigations
generally exclude sample objects with low total attenuation to avoid this problem. 
For instance, \cite{Kriek2013} only used  galaxies with $A_{\rm V}>0.2$
for their study. Here we adopt a lower limit of 
$A_{\rm V}=0.25$ and exclude all spxels with $A_{\rm V}$ below this
limit. Together with the requirement of $S/N>20$, this restriction
gives a sample of 1018 spaxels, which are distributed in 71 galaxies and are used for the analysis in the next section.

\subsection{Global measurements}

For comparison with previous studies, which are mostly limited to
global measurements, we stack all the spaxels within $1\,R_e$ to
generate an integrated spectrum for each of the 134 galaxies in the
SwiM catalog, following the procedure described in \citet{Zhou2020}.
The {\it Swift}/UVOT images and 2MASS Ks-band image are also co-added
correspondingly to derive the photometry within $1\,R_e$ for each
galaxy.  We are left with a sample of 72 galaxies after applying the
same selection criteria: SNR$>20$, and $A_{\rm V}>0.25$. Note that 
the 71 galaxies where the 1018 spaxels come from are all in this sample, 
with one additional galaxy included due to the increase of SNR during 
the stacking. We apply the
same method as described above to perform spectral fitting for each
galaxy and to measure the global dust attenuation properties defined 
in the same way as above. 

\subsection{Tests on mock spectra}
\label{sec:tests}

\begin{figure*}
	\centering
    \includegraphics[width=0.95\textwidth]{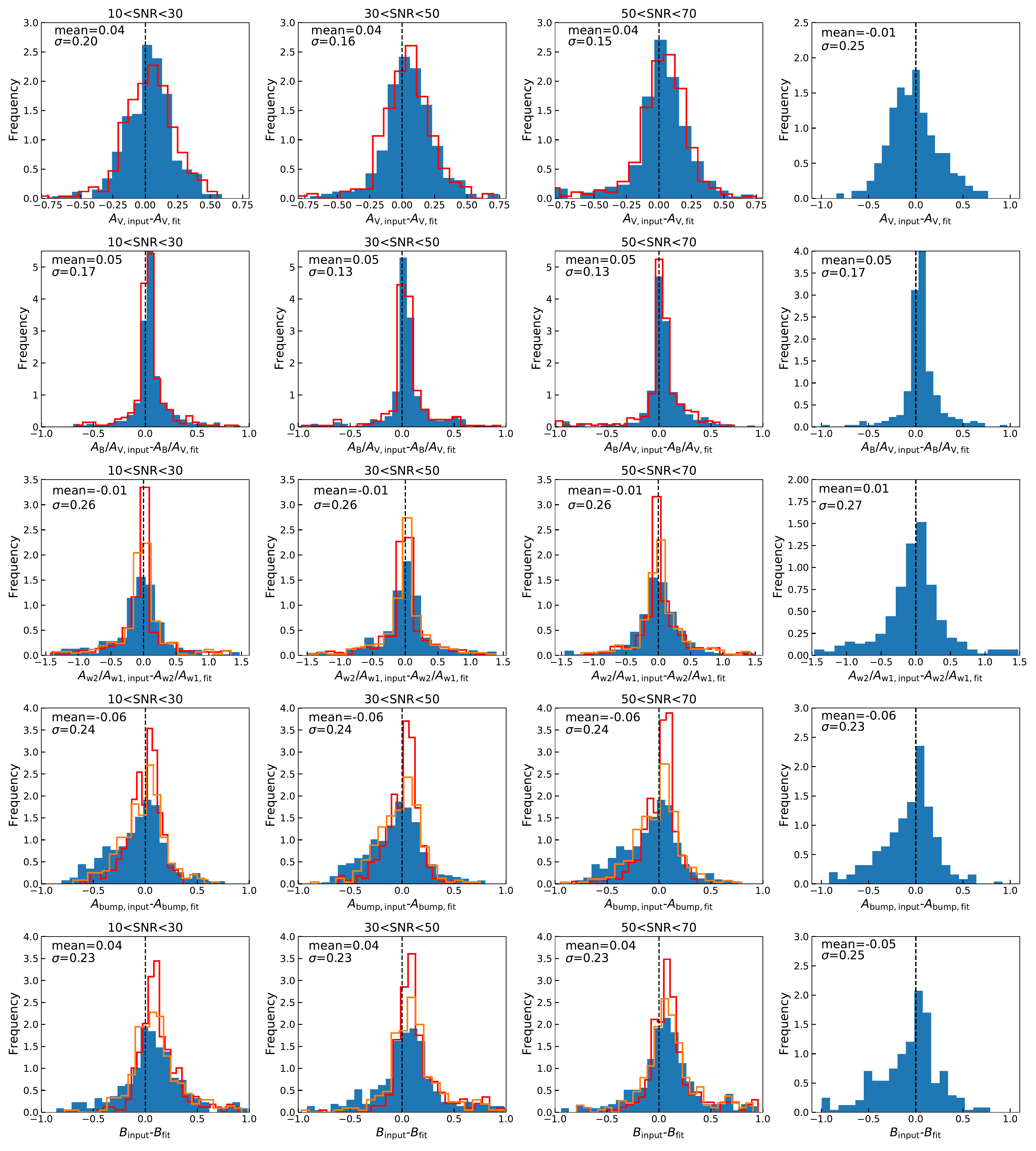}
    \caption{Results of the test of our method on mock spectra. The
      histograms show the distributions of the difference between the
      input and the estimated values for five dust parameters (panels
      from top to bottom): $A_{\rm V}$, $A_{\rm B}/A_{\rm V}$, $A_{\tt
        w2}/A_{\tt w1}$, $A_{\rm bump}$ and $B$.  
      In each row, the left three panels show the results for three SNR 
      ranges (as indicated) as obtained with the BC03 stellar templates,
      while the rightmost panel shows the result with the BPASS stellar
      templates for the SNR range of $10<SNR<30$.
      In each panel, the filled blue histogram shows the result from
      mock spectra that include Gaussian noise with a standard deviation of
      $\sigma_{Ks}=0.07$ in the Ks band and $\sigma_{UV}=0.13$ in the {\tt
        uvw2}, {\tt uvm2} and {\tt uvw1} bands, with the
      mean and standard deviation of the distribution indicated at
      the top-left corner. Red histograms in the top two rows
      are results obtained by assuming a smaller uncertainty in the Ks band
      with $\sigma_{Ks}=0.02$, while the orange and red histograms in the
      lower three rows  show results obtained with smaller
      uncertainties in NUV bands, with $\sigma_{UV}=0.05$ (orange) and
      $\sigma_{UV}=0.02$ (red). All histograms are normalized
      to have a total of unity. Overall, the tests results show that our method 
      adequately recovers the input properties 
      of dust attenuation with systematic and statistical errors well under control.}
    \label{fig:mock}
\end{figure*}

In order to test our method, we have generated a set of mock spectra
that cover a variety of dust attenuation properties and spectral SNRs.
We first obtain various SFHs using the $\Gamma$+B model described in
Section~\ref{sec:method} with randomly generated model parameters.  These
SFHs are combined with the BC03 SSP model spectra assuming a Chabrier
IMF to generate composite spectra covering a wide range in stellar
population properties. Random noise following Gaussian distributions
of different widths is added to the spectra to produce SNRs ranging from
10 to 70. To include dust attenuation with different dust opacity,
curve slopes and 2175{\AA} bump strengths, we use the
functional form proposed by \cite{Conroy2010dust}
(see equations A1-A12 in the paper, hereafter the CSB curve).  
This flexible curve allows variations in $A_{\rm V}$,  slopes 
characterised by $R_{\rm V}$, and the 2175{\AA} bump strength 
characterised by the parameter $B_{\rm CSB}$ (with $B_{\rm CSB}=1.0$ 
being the bump strength of the Milky-Way curve). We randomly generate 
attenuation curves with $0<A_{\rm V}<3$, $1<R_{\rm V}<10$ and  
$0<B_{\rm CSB}<1.5$, and apply them to the mock spectra. 
The dust-attenuated spectra are then convolved with the filter response functions 
of 2MASS Ks and the three {\it Swift}/UVOT bands to calculate the magnitudes. 
We add Gaussian noise with a standard deviation of $\sigma_{Ks}=0.07$ to the Ks 
band and of $\sigma_{UV}=0.13$ to {\tt uvw2}, {\tt uvm2} and {\tt uvw1} bands. 
These values correspond to the typical uncertainties in the real data.  
Repeating this process, we generated a sample of 1000 mock
spectra/SEDs of different SNRs, with known stellar populations and
dust attenuation properties. 

We apply our method to the mock data to estimate the same parameters
as described above and compare them with the input parameters.
\autoref{fig:mock} (the left three columns) shows the results for 
five parameters: $A_{\rm V}$,
$A_{\rm B}/A_{\rm  V}$, $A_{\tt w2}/A_{\tt w1}$,  $A_{\rm bump}$ and $B$.  
Panels from
top to bottom correspond to the five parameters, and panels from left
to right correspond to three ranges of SNRs: $10<$SNR$<30$,
$30<$SNR$<50$, $50<$SNR$<70$, respectively.
In each panel the filled histogram plotted in blue shows the distribution 
of the 1000 mock spectra in
the difference between the input and output parameters.  
It is encouraging that, overall, all
the four parameters are well recovered with only weak or no bias. 
For the dust parameters in the optical,
i.e. $A_{\rm V}$ and $A_{\rm B}/A_{\rm V}$, the standard deviation
$\sigma$ decreases significantly as one goes from the first SNR bin,
where $\sigma(A_{\rm V})=0.20$ and $\sigma(A_{\rm B}/A_{\rm V})=0.17$,
to the second SNR bin, where $\sigma(A_{\rm V})=0.16$ and
$\sigma(A_{\rm B}/A_{\rm V})=0.13$, with no further decrease
as the SNR exceeds 30.  For the parameters in the UV, i.e.  $A_{\rm
  w2}/A_{\rm w1}$, $A_{\rm bump}$ and $B$, the standard deviation does not depend on
SNR, with $\sigma(A_{\rm w2}/A_{\rm w1})=0.26$, $\sigma(A_{\rm bump})=0.24$ and 
$\sigma(B)=0.23$ at all SNRs.

Using mock spectra generated in a similar way, \citet{Li2020} found that
their method can reproduce $E(B-V)$ without bias but with a
standard derivation of $\sigma\sim0.022-0.05$ mag at $10<$SNR$<30$
(see their Fig.7), corresponding to $\sim0.1-0.2$ mag in $A_{\rm V}$
assuming a Calzetti curve with $R_{\rm V}=4.05$. For the same SNR
range we find a similar but slightly larger standard deviation with
$\sigma(A_{\rm V})=0.20$ mag, implying that the uncertainties in the
dust parameters are not purely limited to the spectral SNR.  The
uncertainty in the Ks-band photometry could be one of the
factors. We have repeated the same test but adopting a smaller
uncertainty in the Ks-band magnitude: $\sigma_{\rm Ks}=0.02$, and we show
the result as red histograms in the upper two rows in
\autoref{fig:mock}. The results are similar to the filled blue histograms
obtained with $\sigma_{\rm Ks}=0.07$, suggesting that the NIR photometry
is not a dominating source of uncertainty in the estimated dust parameters. 
Similarly, we have done tests using mock data with smaller uncertainties 
in the UV photometry, $\sigma_{\rm UV}=0.05$ and 0.02 for the 
three {\it Swift}/UVOT bands.  The predicted distributions,  
plotted as orange and red histograms in the lower three rows
in \autoref{fig:mock}, appear to be significantly 
narrower than the corresponding filled blue histograms. 
Together with the weak dependence on SNR, this indicates that the 
uncertainties in $A_{\rm w2}/A_{\rm w1}$, $A_{\rm bump}$ and $B$ are dominated 
by the {\it Swift}/UVOT photometry.

The results presented above are based on the stellar population model of BC03. 
Although this model has been widely used, 
different stellar population templates \citep[e.g.][]{Maraston2005,Vazdekis2010,Eldridge2017} may lead to different 
results, particularly in NIR and UV. To fully account for the
systematics introduced by different SSP templates is beyond the
scope of this work. Here we present a simple test to examine the impacts 
of this issue. In this test, we apply the same method to the same set of mock spectra generated from BC03 SSP templates, but the SSP models from the Binary Population and Spectral 
Synthesis \citep[BPASS;][]{Eldridge2017} instead of the BC03 models are used in the fitting process.
Compared to the BC03 models, the BPASS templates take into account the evolution of 
binaries, and thus are different in many aspects, especially in the UV range.
The results of this test  are shown in the rightmost column in \autoref{fig:mock},
but only for the lowest S/N range ($10<$SNR$<30$) for simplicity. 
Comparing the results with the blue filled
histograms shown in the leftmost column, we see that
the standard deviation of \Av~increases slightly from 0.2 to 0.25 mag, 
but that all the five parameters are recovered with similarly weak bias 
and small standard deviations. These results indicate that the uncertainties 
in dust slopes and bump strengths are mostly dominated by uncertainties 
in flux measurements and the method itself, and that using different SSP 
templates should not significantly affect our main results.

\begin{figure*}
	\centering
	\includegraphics[width=1.0\textwidth]{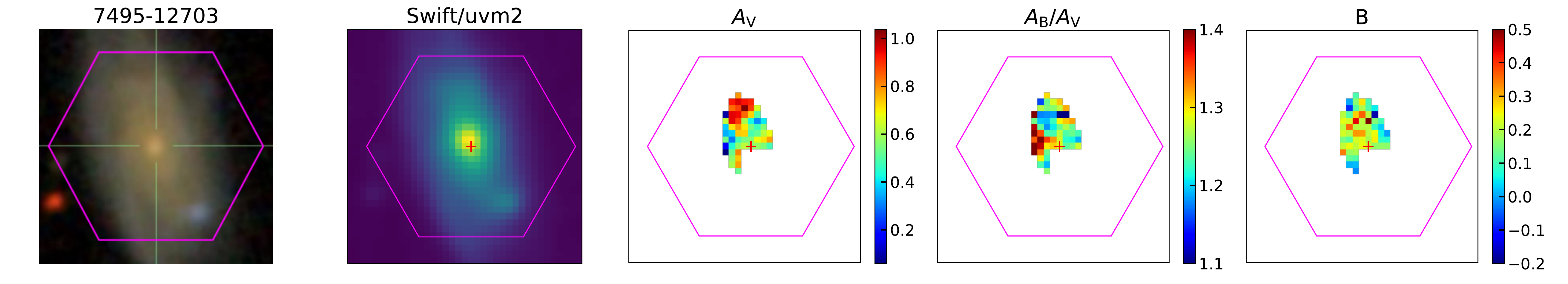}	
	\includegraphics[width=1.0\textwidth]{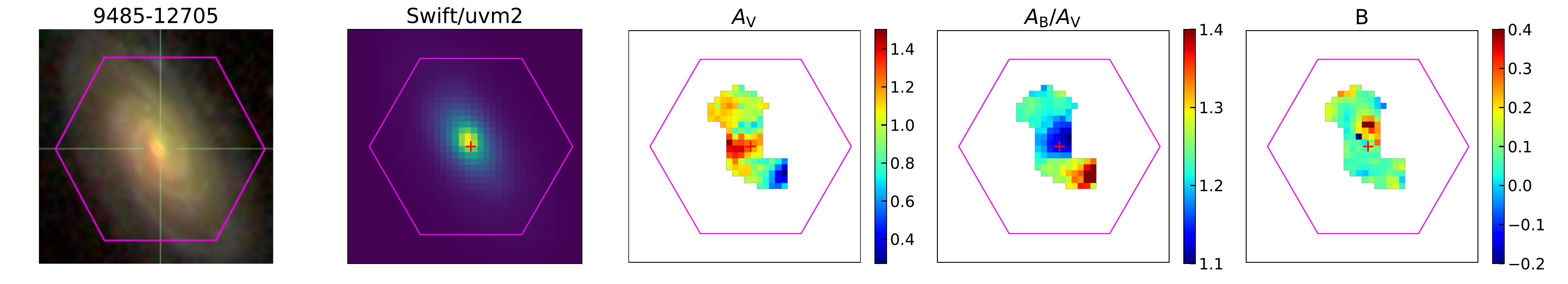}
	\caption{Two example galaxies in our sample. From left to right, plots show the optical image, Swift {\tt uvm2} image, the absolute attenuation in V band (\Av),  the slope of the attenuation curve in optical ($A_{\rm B}/A_{\rm V}$), and the bump strength $B$ in each galaxy. In each panel, the MaNGA footprint is shown in magenta, with a red cross marking the center of the integral field.}
	\label{fig:example}
\end{figure*}

In the tests above, the uncertainties in the spectra and SEDs are 
modeled to some degree by including a component of Gaussian noise. 
However, real observational uncertainties may be more complicated.  
For example, uncertainties in flux calibration and in sky contamination
may not be characterised by a simple Gaussian noise and may lead to
systematic bias. In addition, it is possible that a small population of young or intermediate-age stars in the galaxy that do not affect the optical spectrum very much may contribute to the light in NUV. In this case the difference between the best-fit and observed {\tt uvw2}, {\tt uvm2} and {\tt uvw1} fluxes will not give an accurate estimate of $A_{\tt w2}$, $A_{\tt m2}$, $A_{\tt w1}$. However, the $\Gamma$+B model used in our mock test has naturally allowed the presence of an arbitrary fraction of stars of different ages and metallicities. The similarity of the overall scatter as seen in the mock and real spectral fitting (see results in
\autoref{sec:results}) indicates that these effects are not significant and 
the main uncertainties should have been
captured in our method.

Based on the tests we conclude  that, as long as the spectral SNR is
sufficiently high (e.g. SNR$>10$ considered in the test), the NIR
photometry is not the main source of uncertainties. The uncertainties 
in the dust parameters in UV may be further reduced by increasing the 
quality of the UV photometry. Given the quality of the real data
used in our study, SNR$>20$, $\sigma_{\rm Ks}\sim0.07$ and
$\sigma_{\rm UV}\sim0.13$, our method is expected to recover
the average properties of dust attenuation in individual spaxels
with systematic and statistical errors well under control. 

\section{Results}
\label{sec:results}

We obtained the dust properties in our sample galaxies using the method presented above. Before digging into detailed statistical results for the dust parameters, we show in \autoref{fig:example} two example galaxies that have a large number of spaxels satisfying our selection criteria. As seen from the plot, the dust properties, including the absolute attenuation in V band (\Av),  the slope of the attenuation curve in optical ($A_{\rm B}/A_{\rm V}$), and the 2175{\AA} bump strength $B$ , vary significantly across the galaxy, which signifies the importance of spatially resolve data such as MaGNA and Swift/UVOT in investigating the dust properties in galaxies. It is also noticeable that these dust properties are not randomly distributed in the galaxy, but rather some correlations are seen between the parameters. In what follows we will use the entire sample to examine the variation of the different dust properties, as well as the correlations between the dust properties.

\subsection{Amplitudes and slopes of the attenuation curves}

\begin{figure*}
	\centering
	\includegraphics[width=0.48\textwidth]{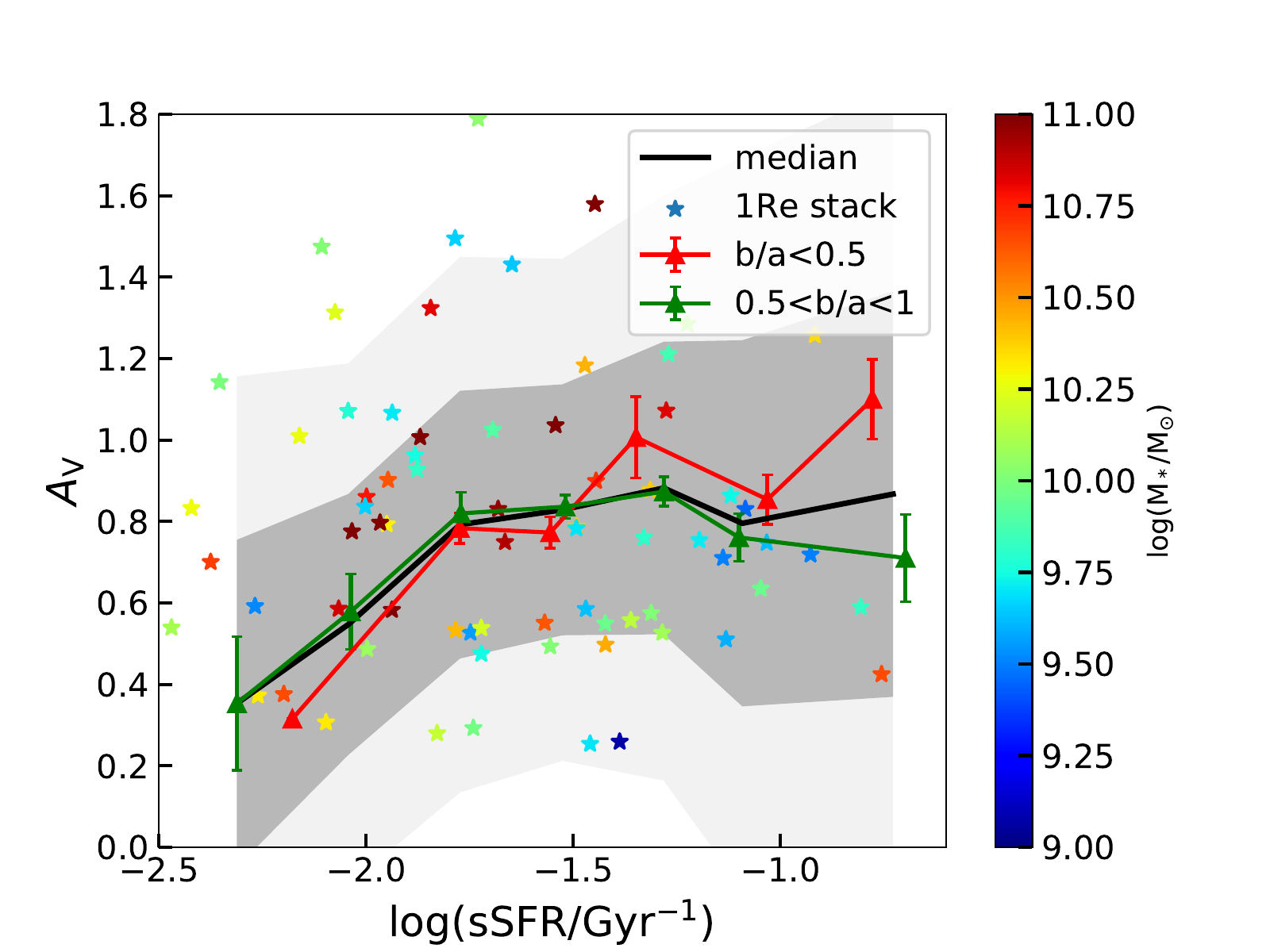}	
	\includegraphics[width=0.48\textwidth]{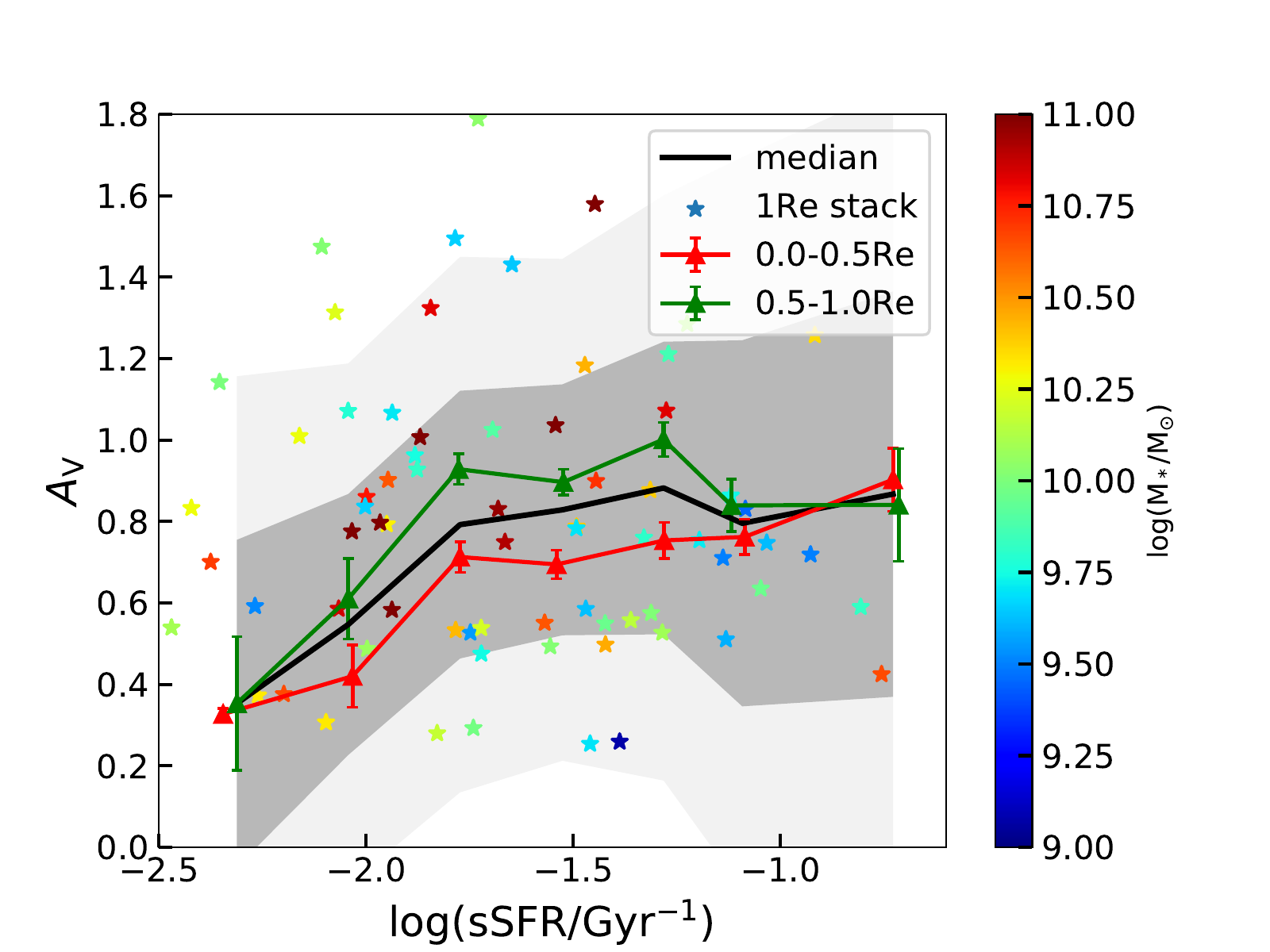}
	\caption{{\it Left:} The absolute attenuation in V band (\Av) as a function of the sSFR. The black line shows the median relation of all the pixels in our galaxies, and the grey and light-grey shaded regions indicate the 1$\sigma$ and 2$\sigma$ scatter of individual spaxels around the median. Stars are the results from the stacked spectra within $1R_e$ of each galaxy, color-coded by the total stellar mass of the galaxy indicated by the color bar. The red/green triangles/lines show the results for subsamples of spaxels selected by $b/a$, the minor-to-major axis ratio of the SDSS $r$-band image of the host galaxies. {\it Right:} Same as left, except that the red/green triangles/lines are for subsamples selected by the galactocentric distance of the spaxels.}
	\label{fig:AV_sSFR}
\end{figure*}

We first examine the amplitude and slope of the dust attenuation curve
and their dependence on star formation rate and stellar mass.  To this end, we
take the measurements of H$\alpha$ emission line flux provided by the
MaNGA DAP for all the spaxels in our sample, and estimate a star
formation rate (SFR) for each spaxel from the H$\alpha$ luminosity by  $\rm
SFR(M_{\odot}yr^{-1})=7.9\times10^{-42}L(H\alpha)$
\citep{Kennicutt1998ARA}. Stellar masses for individual spaxels are obtained from the MaNGA 
spectra as a byproduct of the technique of \cite{Li2020} described in 
\S~\ref{sec:attenuation_curve}. We have also obtained the global measurements of 
the SFR for the 72 galaxies by adding up SFRs for all individual spaxels in each galaxy, 
while the total stellar masses of these galaxies are taken from the NSA.

In \autoref{fig:AV_sSFR} we examine the correlations of the absolute attenuation
in the V band (\Av)  with the specific star formation rate (sSFR) on a 
logarithmic scale, $\log_{10}({\rm sSFR})\equiv\log_{10}({\rm SFR}/M_\ast)$.  In both panels,
the solid black line displays the median of all the spaxels, while the gray 
and light gray regions indicate the 1$\sigma$ and 2$\sigma$ scatter of 
individual spaxels around the median. The global measurements of the 
72 galaxies are plotted as colored stars, color-coded by their total stellar
mass. As can be seen, our samples of both individual 
spaxels and the global measurements span a wide range in both \Av\ and
$\log_{10}({\rm sSFR})$. The global measurements present a similar
distribution to the spaxels, and the distribution does not depend 
significantly on the total stellar mass of galaxies. The median \Av\ of spaxels 
increases from \Av$\sim0.3$ mag at the lowest sSFR to \Av$\sim0.8$ mag at
$\log_{10}({\rm sSFR}/{\rm Gyr}^{-1})\sim-1.8$, remaining roughly at
a constant level at higher sSFRs. This is consistent with that obtained by
\citet{Li2021} who found a similar trend of the stellar color excess
$E(B-V)$ of kpc-sized regions with sSFR (see figure 14 in that paper).
The scatter among spaxels at a given sSFR is $\sigma\sim 0.4$ mag, and
shows a slightly positive correlation with sSFR. Assuming an average
scatter of $\lesssim$0.2 mag expected to be caused by the
uncertainties in our method and data (see \S~\ref{sec:tests}) and a
total scatter of $\sim$0.4 mag as shown in the figure, we estimate
that on average the intrinsic scatter is $\sigma\ga 0.35$ mag for our
sample. 

\begin{figure*}
	\centering
	\includegraphics[width=0.32\textwidth,height=0.25\textwidth]{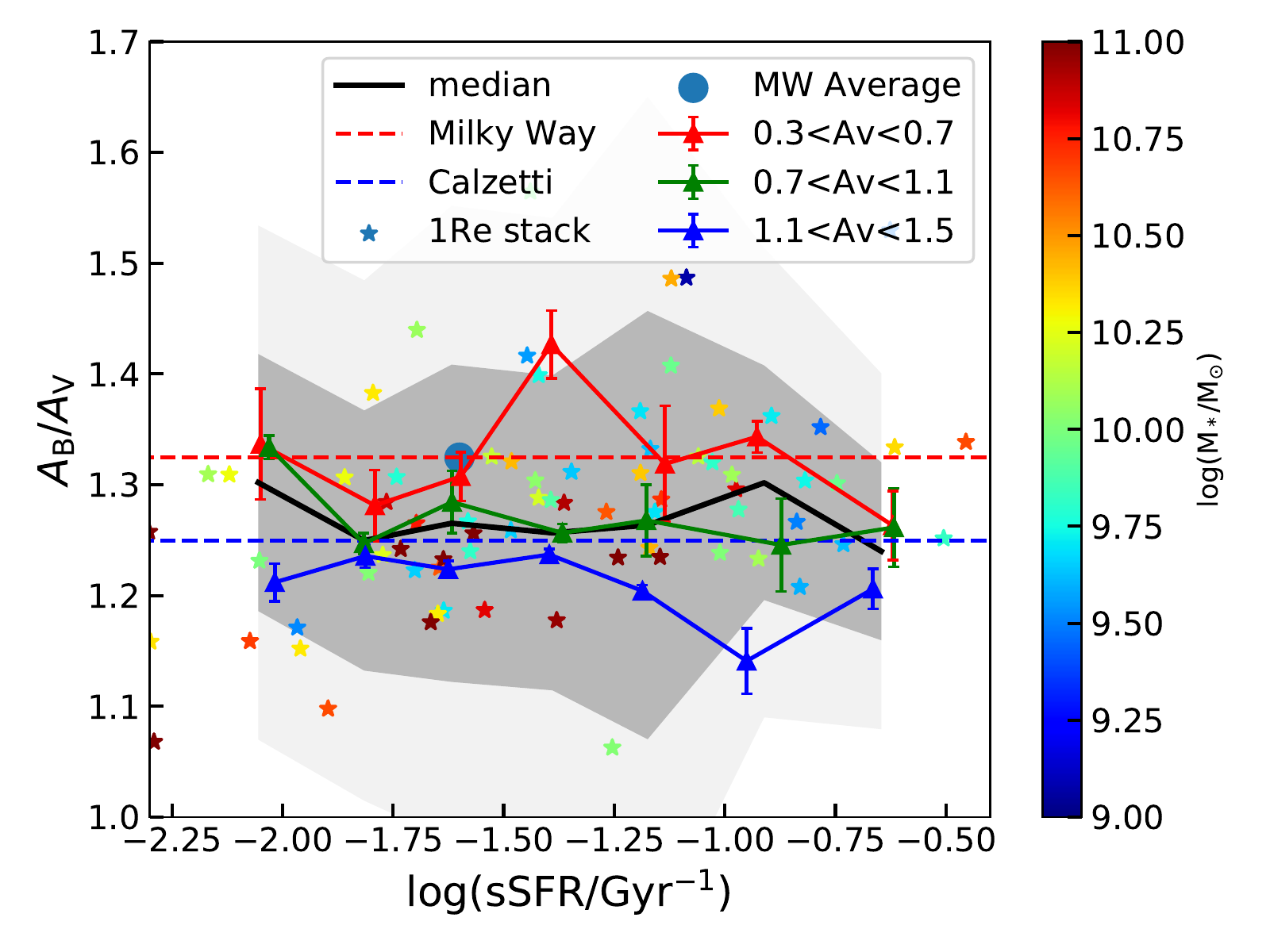}
	\includegraphics[width=0.32\textwidth,height=0.25\textwidth]{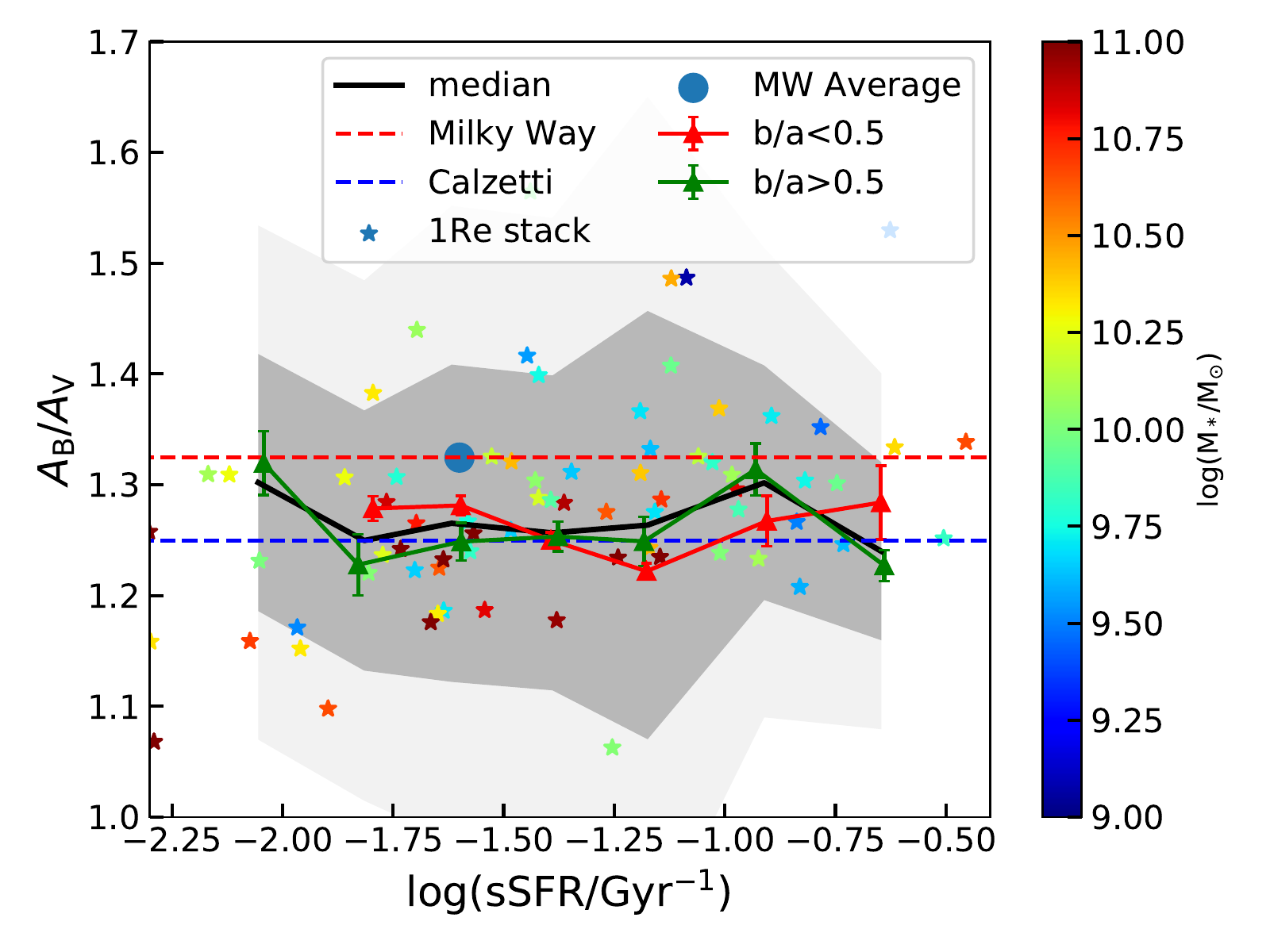}
	\includegraphics[width=0.32\textwidth,height=0.25\textwidth]{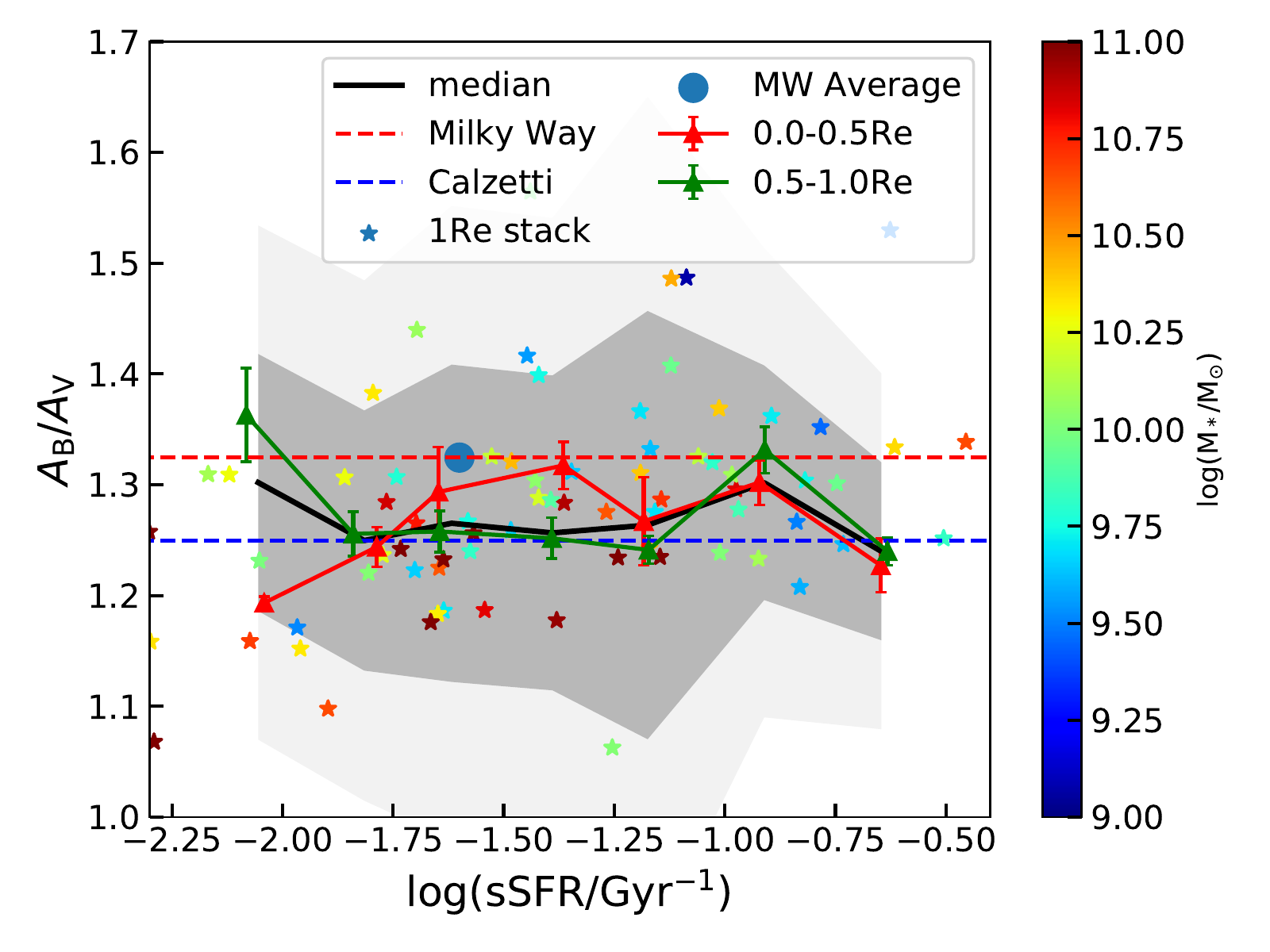}
	
	\caption{The slope of the attenuation curve in optical ($A_{\rm B}/A_{\rm V}$) as a function of the sSFR. The black line shows the median values of the results obtained from individual pixels, with grey shaded regions indicates 1$\sigma$ and 2$\sigma$ scatters of the data points. Stars are results from the 1Re stacks of each galaxy, with color codes showing the total stellar mass of the galaxy (from NSA). In panels from left to right, colored triangles/lines show results of subsamples of spaxels selected by \Av\ (left), $b/a$ (middle) and the galactocentric distance (right), respectively.}
	\label{fig:slope_opt_sSFR}
\end{figure*}

\begin{figure*}
	\centering
	\includegraphics[width=0.32\textwidth,height=0.25\textwidth]{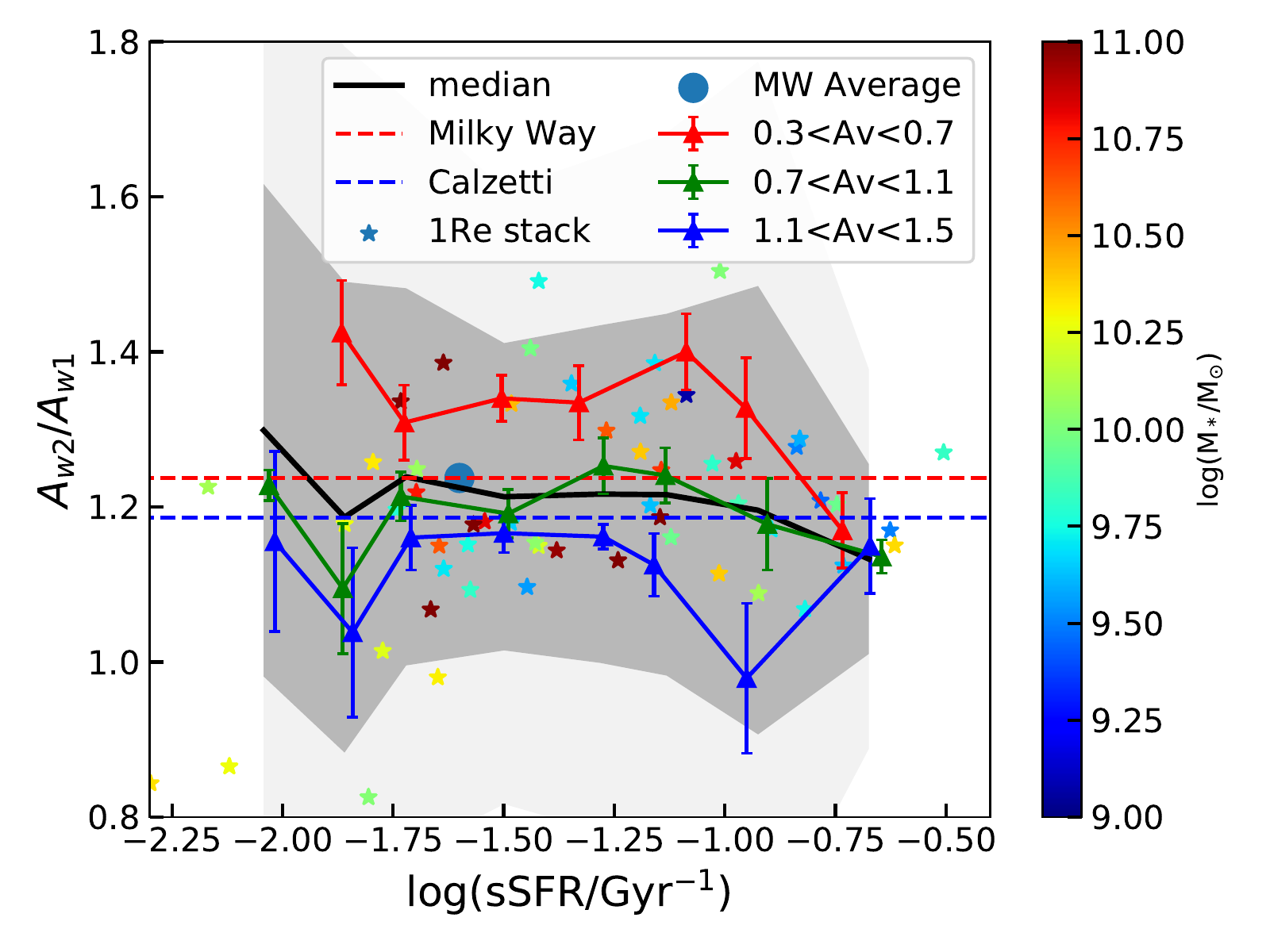}
	\includegraphics[width=0.32\textwidth,height=0.25\textwidth]{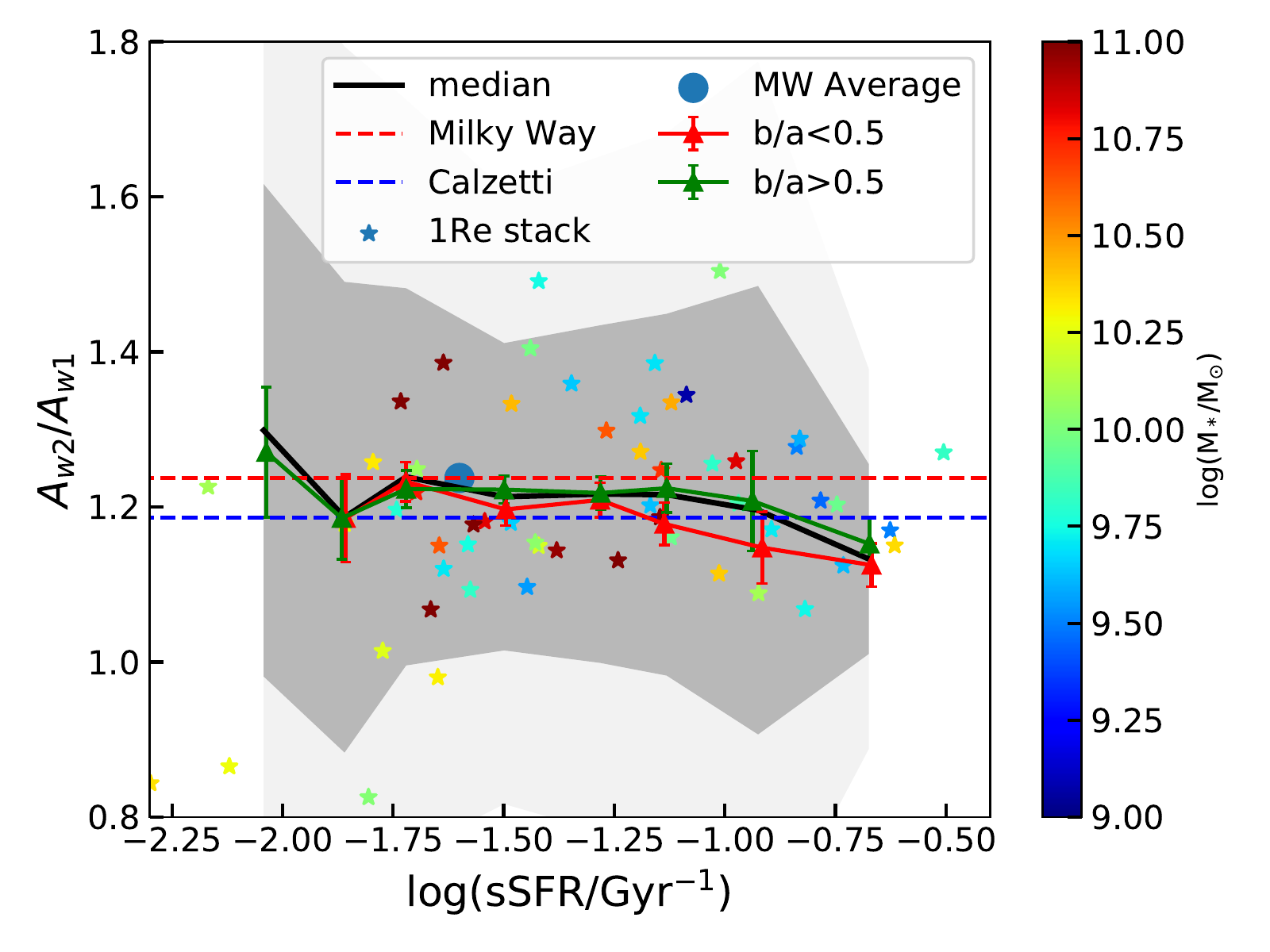}
	\includegraphics[width=0.32\textwidth,height=0.25\textwidth]{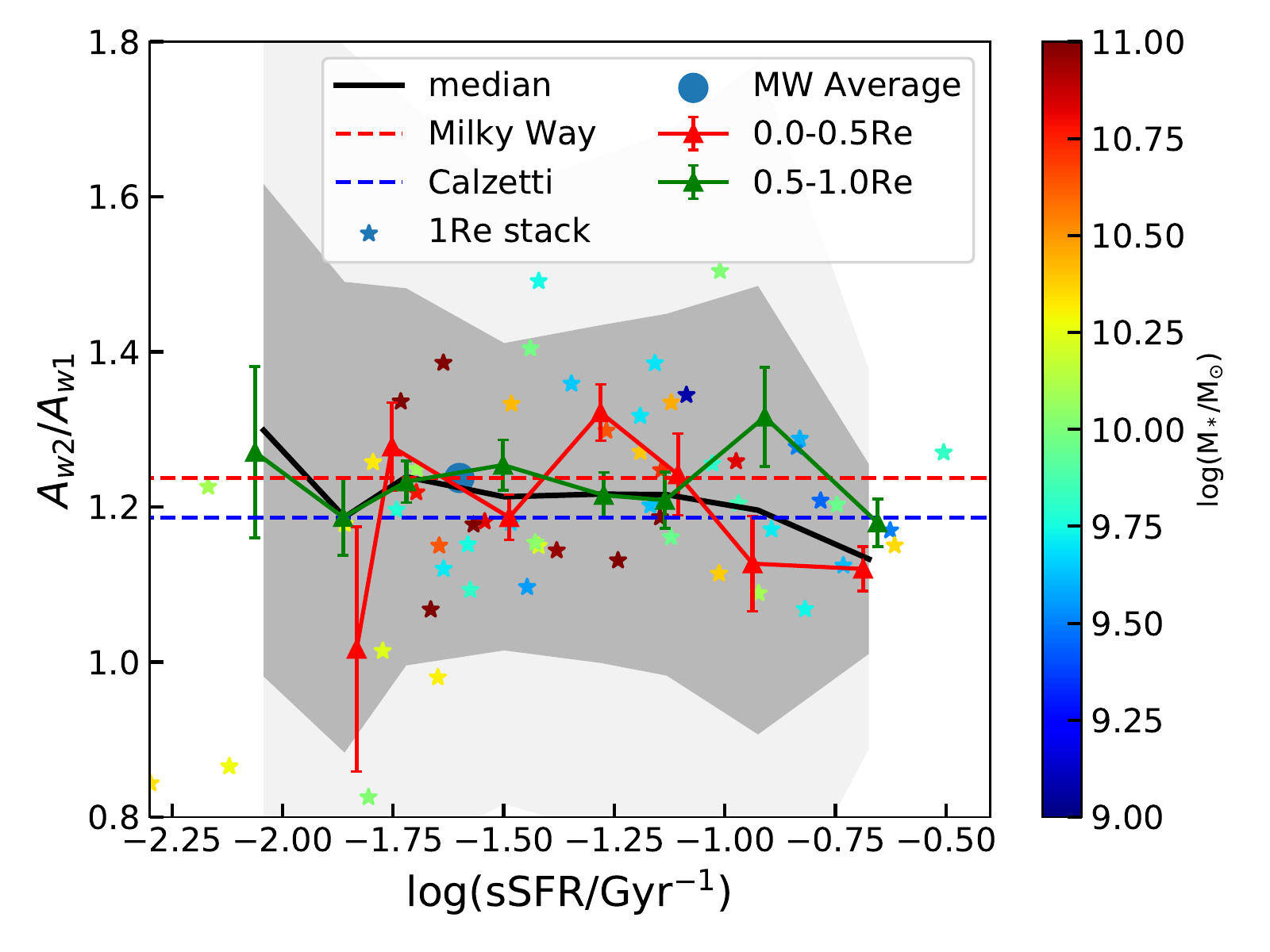}
	
	\caption{The slope of the attenuation curve in NUV ($A_{\tt w2}/A_{\tt w1}$). The black line shows the median values of the results obtained from individual pixels, with grey shaded regions indicates 1$\sigma$ and 2$\sigma$ scatters of the data points. Stars are results from the 1Re stacks of each galaxy, with color codes showing the total stellar mass of the galaxy (from NSA). In panels from left to right, colored triangles/lines show results of subsamples of spaxels selected by \Av\ (left), $b/a$ (middle) and the galactocentric distance (right), respectively.}
	\label{fig:slope_NUV_sSFR}
\end{figure*}

\begin{figure*}
	\includegraphics[width=0.33\textwidth,height=0.25\textwidth]{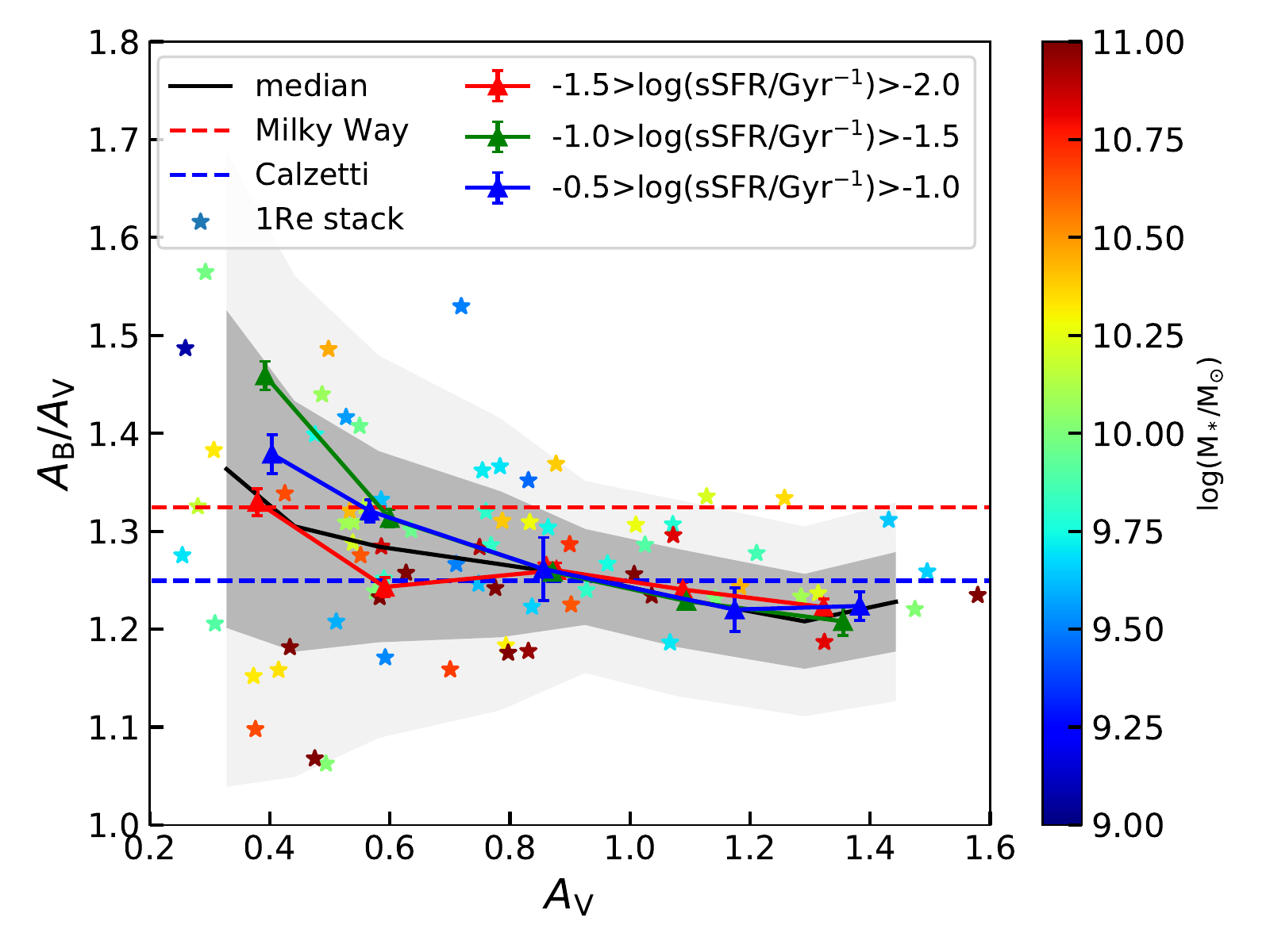}
	\includegraphics[width=0.33\textwidth,height=0.25\textwidth]{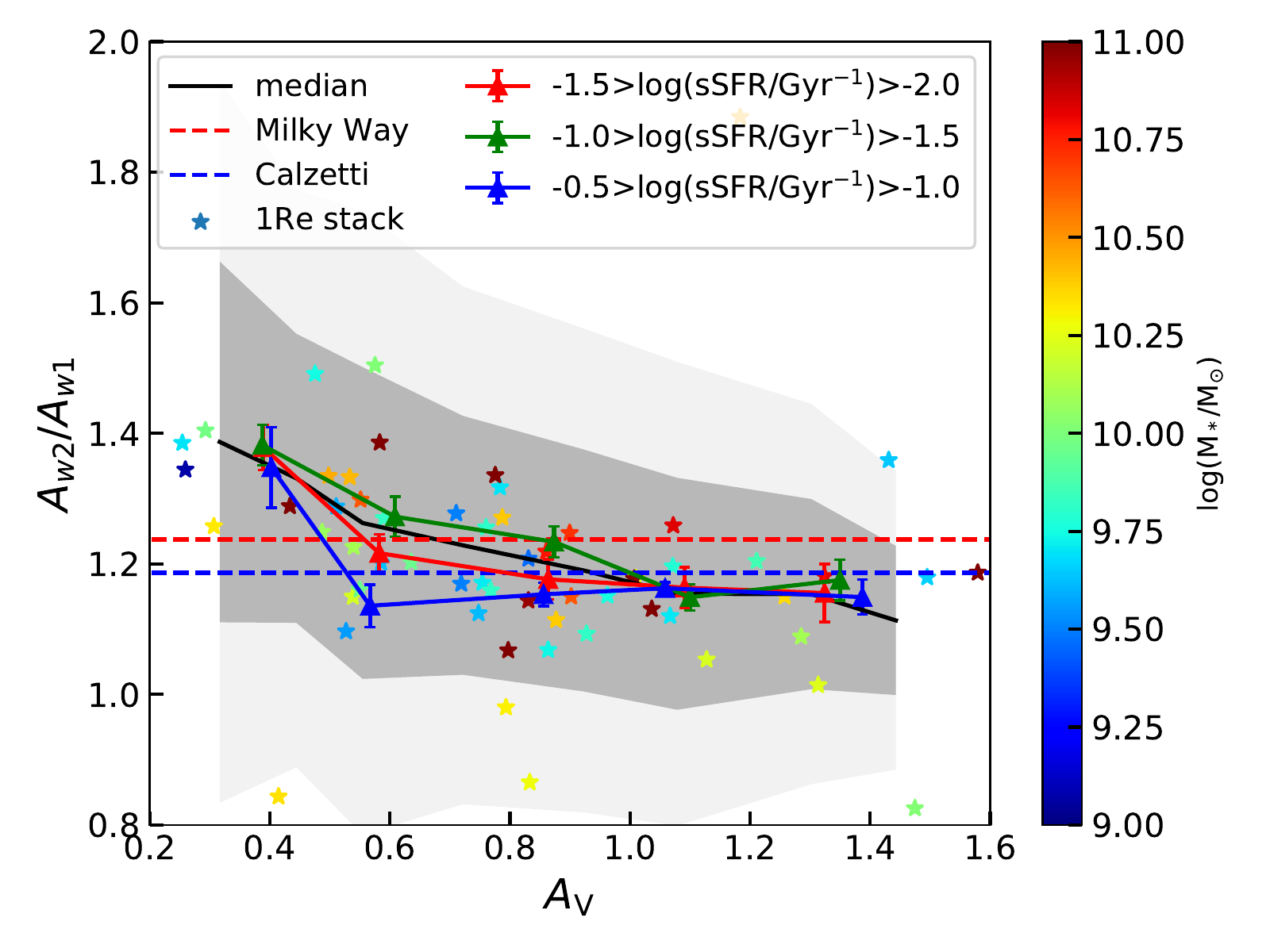}
	\includegraphics[width=0.33\textwidth,height=0.25\textwidth]{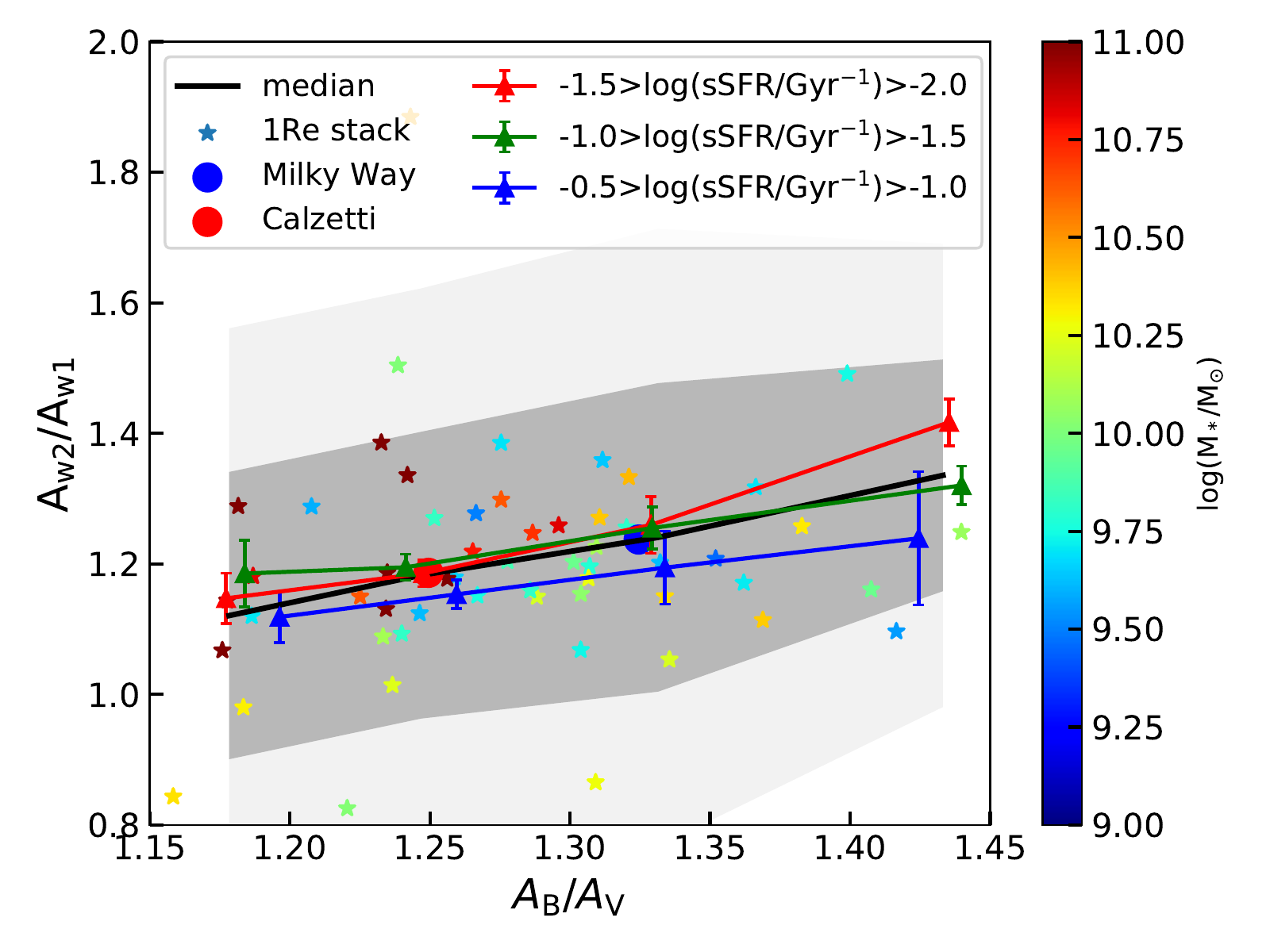}
	\caption{The mutual relation between the dust opacity in V-band (\Av) and the slopes of the attenuation curve  in optical ($A_{\rm B}/A_{\rm V}$, left) and NUV ($A_{\tt w2}/A_{\tt w1}$, middle), with the mutual relation between $A_{\rm B}/A_{\rm V}$ and $A_{\tt w2}/A_{\tt w1}$ showing in the right panel. The black line shows the median values of the results obtained from individual pixels, with grey shaded regions indicates 1$\sigma$ and 2$\sigma$ scatters of the data points. Stars are results from the 1Re stacks of each galaxy, with color codes showing the total stellar mass of the galaxy (from NSA). In the left and middle panels the slopes for the Milky Way curve and Calzetti curve are indicated as blue and red dash lines, while in the right panel they are marked as blue and red dots respectively.}
	\label{fig:curve_relation_AV}
\end{figure*}

We have attempted to examine the potential sources responsible 
for the intrinsic scatter in the $A_V$-sSFR relation by dividing the 
spaxels of given sSFR into two subsamples according to either the 
minor-to-major axis ratio ($b/a$) of the SDSS $r$-band image of 
the host galaxy or the galactocentric distance of the spaxels scaled by 
the effective radius ($R_e$) of the galaxy. The results of the 
subsamples selected by $b/a$ are plotted in the left panel of 
\autoref{fig:AV_sSFR} as red/green triangles connected by 
red/green lines. We find no dependence on $b/a$, indicating that galaxy inclination (geometry of 
dust distribution) does not explain the observed scatter. The results of 
the subsamples selected by the galactocentric distance of spaxels 
are shown in the right panel of the same figure. For comparison, 
the results of the full samples of both spaxels and galaxies 
shown in the left panel are repeated here. 
A weak but significant dependence 
on the galactocentric distance is seen at intermediate sSFRs, where 
spaxels within $0.5R_e$ tend to have smaller $A_V$ at fixed sSFR 
than spaxels located between $0.5R_e$ and $R_e$.
However, the difference in $A_V$ is at a level of $0.2$ mag at most 
between the two subsamples, implying that the location of spaxels 
cannot fully account for the overall scatter. The residual dependence 
on the location of spaxels indicates some variation of dust
attenuation between the central bulge and the outer disk, but 
more work is needed to better understand the variation.

Next,  we examine the correlation of the attenuation curve slope in 
optical ($A_{\rm B}/A_{\rm V}$) with the sSFR, and the results are shown  
in \autoref{fig:slope_opt_sSFR}. The median of and scatter among 
spaxels are plotted as the black solid line and shaded region, respectively,  
and the global measurements are plotted as colored stars. 
In panels from left to right, colored triangles/lines 
show results of subsamples of spaxels selected by 
\Av\ (left), $b/a$ (middle) and the galactocentric distance (right), respectively. 
In each panel, the blue dashed line indicates the Calzetti curve, 
which has $A_{\rm B}/A_{\rm V}=1.25$, while the red dashed line indicates
the Milky Way-type curve with $A_{\rm B}/A_{\rm V}=1.32$. 
The Milky Way itself is plotted as a big blue dot in each panel. Overall, 
$A_{\rm  B}/A_{\rm V}$ presents a median
value that is roughly constant at all sSFRs and is very close to the
slope of the Calzetti curve, with a 1$\sigma$ scatter of
$\sigma(A_{\rm  B}/A_{\rm V})\sim 0.15$ quite independent
of the sSFR. The Milky Way-type curve falls within the 1$\sigma$ range.  

In \autoref{fig:slope_NUV_sSFR} we further examine the correlation of 
the slope in UV ($A_{\tt w2}/A_{\tt w1}$) with sSFR. Symbols/lines/colors
are coded in the same way as in the previous figure. We find 
that $A_{\tt w2}/A_{\tt w1}$ appears to have an anti-correlation with 
sSFR, but the correlation is quite weak given the large scatter among the spaxels.
The median value of $A_{\tt w2}/A_{\tt w1}$ is closer to the Milky-Way type at low sSFR and
closer to the Calzetti curve at high sSFR, despite the small
difference between the two model curves. This indicates the flattening
of the NUV attenuation curve in star-forming regions. 

In both \autoref{fig:slope_opt_sSFR} and  \autoref{fig:slope_NUV_sSFR}
we do not observe any clear trends with the total stellar mass of
galaxies. It is interesting to note that the 1$\sigma$ scatter in
both figures, $\sigma\sim0.15$ for $A_{\rm  B}/A_{\rm V}$ and
$\sigma\sim0.25$ for $A_{\tt w2}/A_{\tt w1}$, is comparable to  the
scatter predicted by the test in \S~\ref{sec:tests},
$\sigma\lesssim 0.17$ and $\sigma=0.26$ for the two parameters, 
respectively. This indicates that the intrinsic scatter in both $A_{\rm
	B}/A_{\rm V}$ and $A_{\tt w2}/A_{\tt w1}$
	is small once the sSFR is limited
to a narrow range. When dividing spaxels into subsamples, we find 
no significant dependence of the slope-sSFR relation on $b/a$ and the 
galactocentric distance, suggesting that the inclination of the host 
galaxy and the location within the host are not the reasons for the 
scatter. In contrast, at fixed sSFR the slopes in 
both optical and NUV show significant and systematic trends with 
\Av. This effect is more pronounced in NUV where the dependence on 
\Av\ is similarly seen for all sSFRS except at the highest end. 
For the optical slope, the \Av\ dependence is mainly seen in the intermediate 
range of sSFR. 

\begin{figure*}
	\includegraphics[width=0.33\textwidth,height=0.25\textwidth]{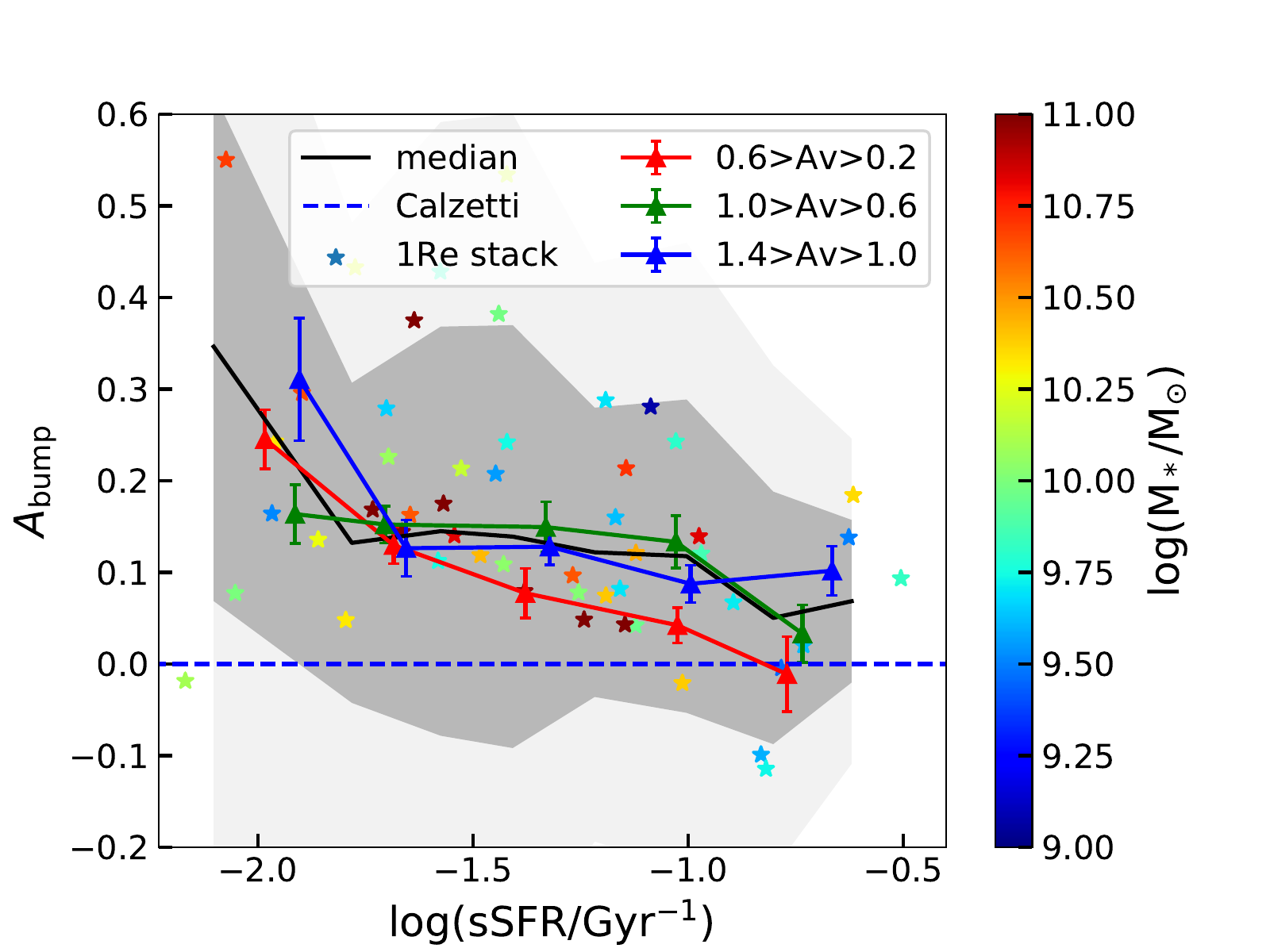}
	\includegraphics[width=0.33\textwidth,height=0.25\textwidth]{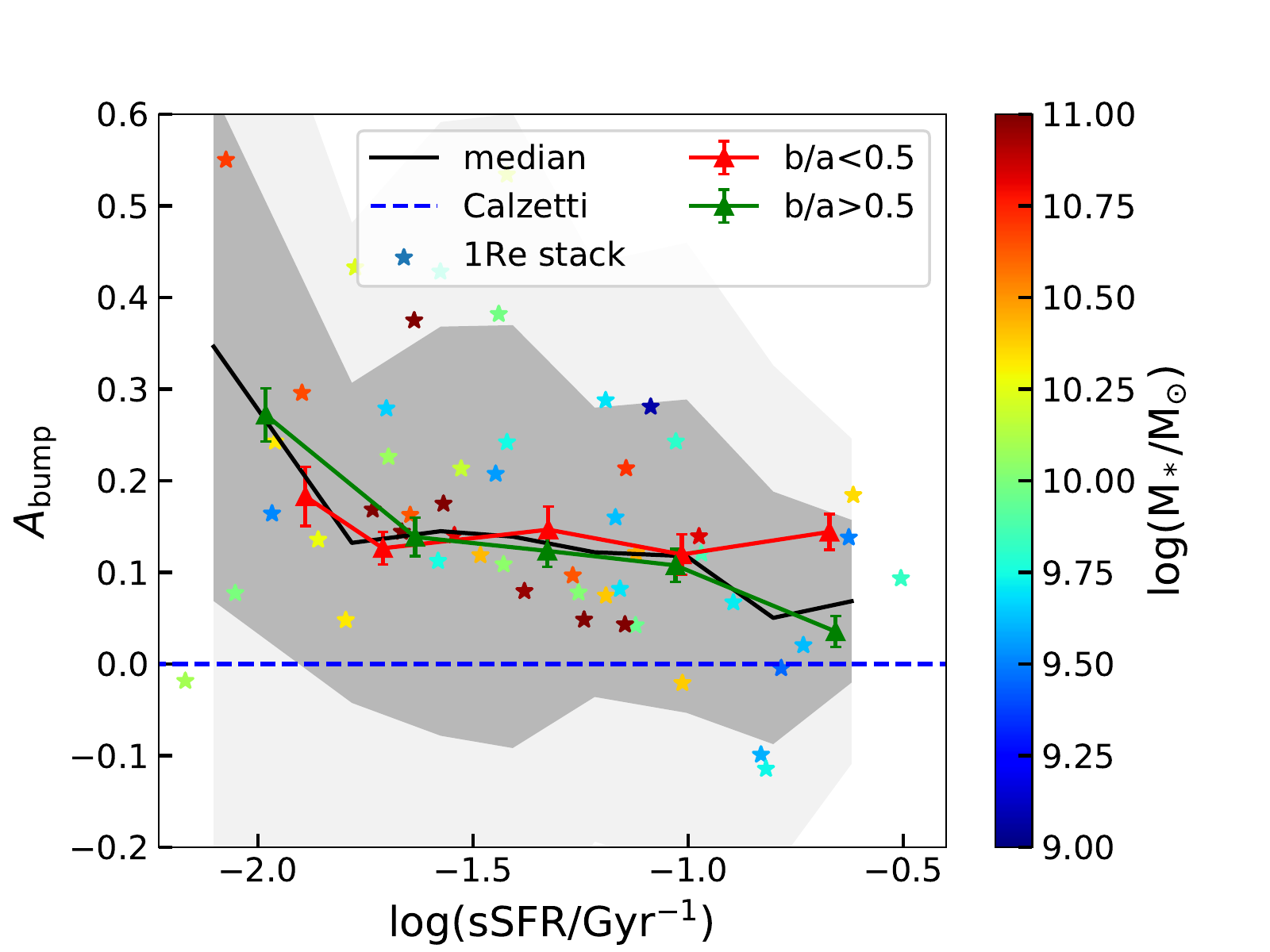}
	\includegraphics[width=0.33\textwidth,height=0.25\textwidth]{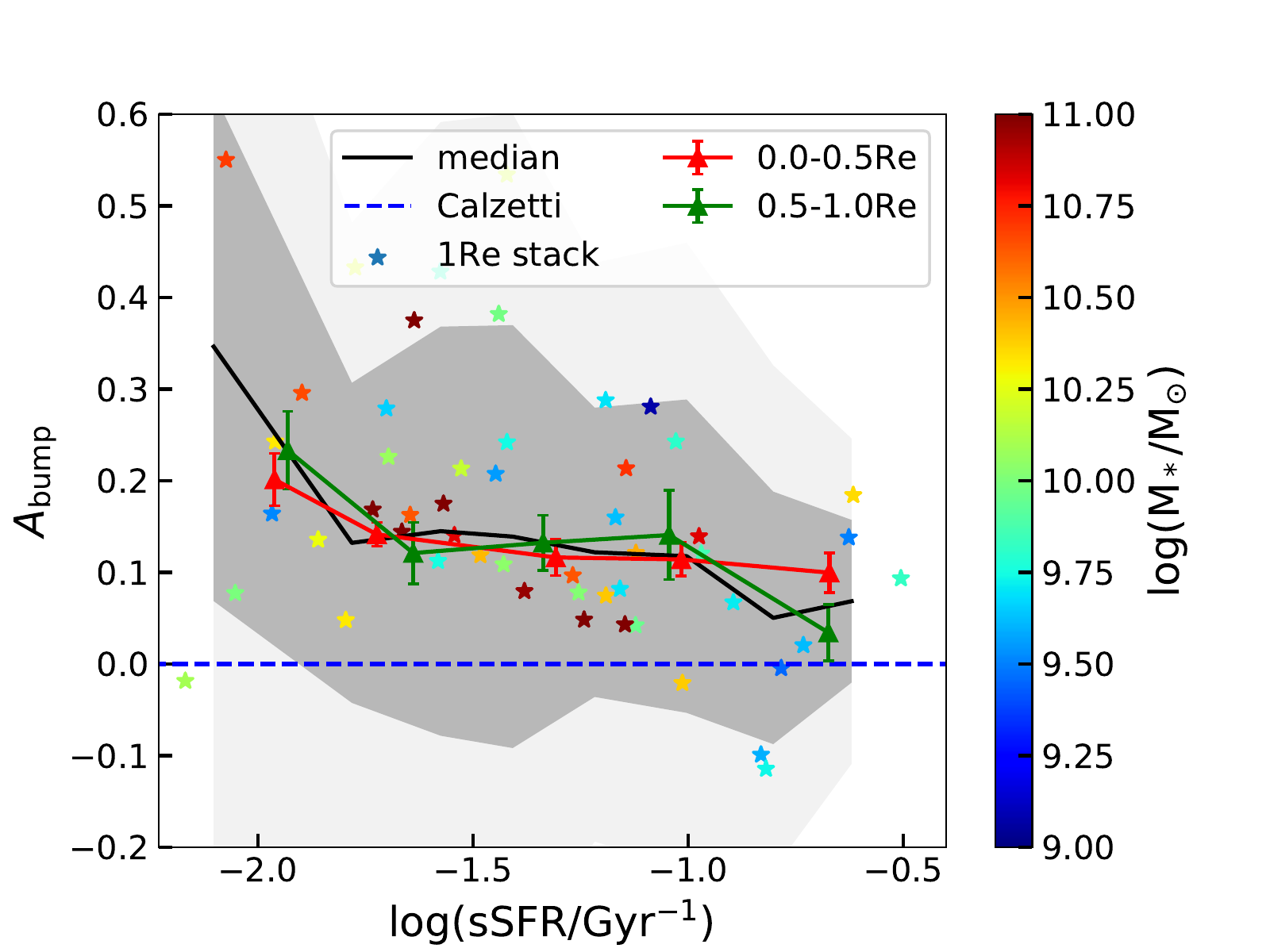}\\
	\includegraphics[width=0.33\textwidth,height=0.25\textwidth]{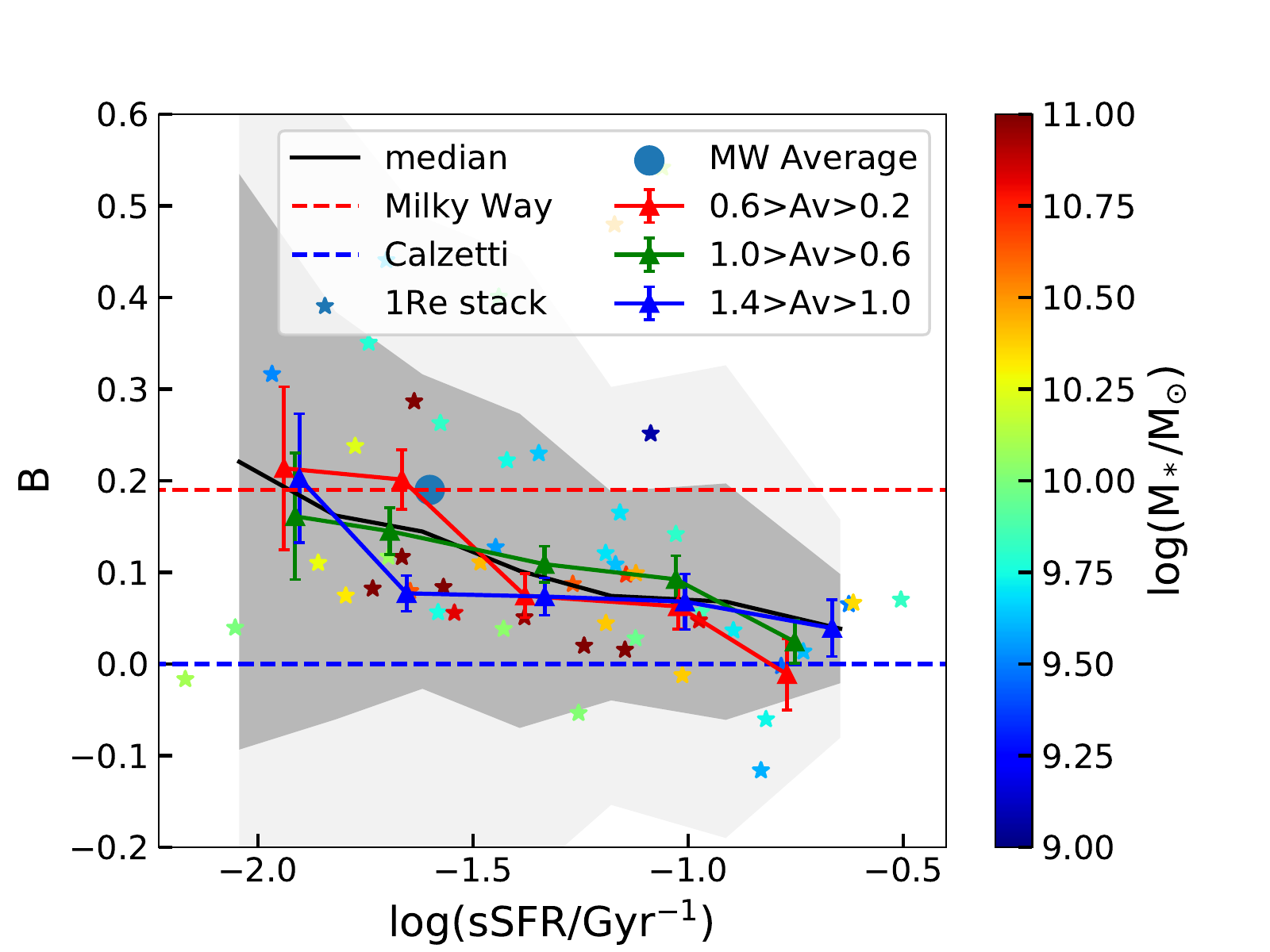}
	\includegraphics[width=0.33\textwidth,height=0.25\textwidth]{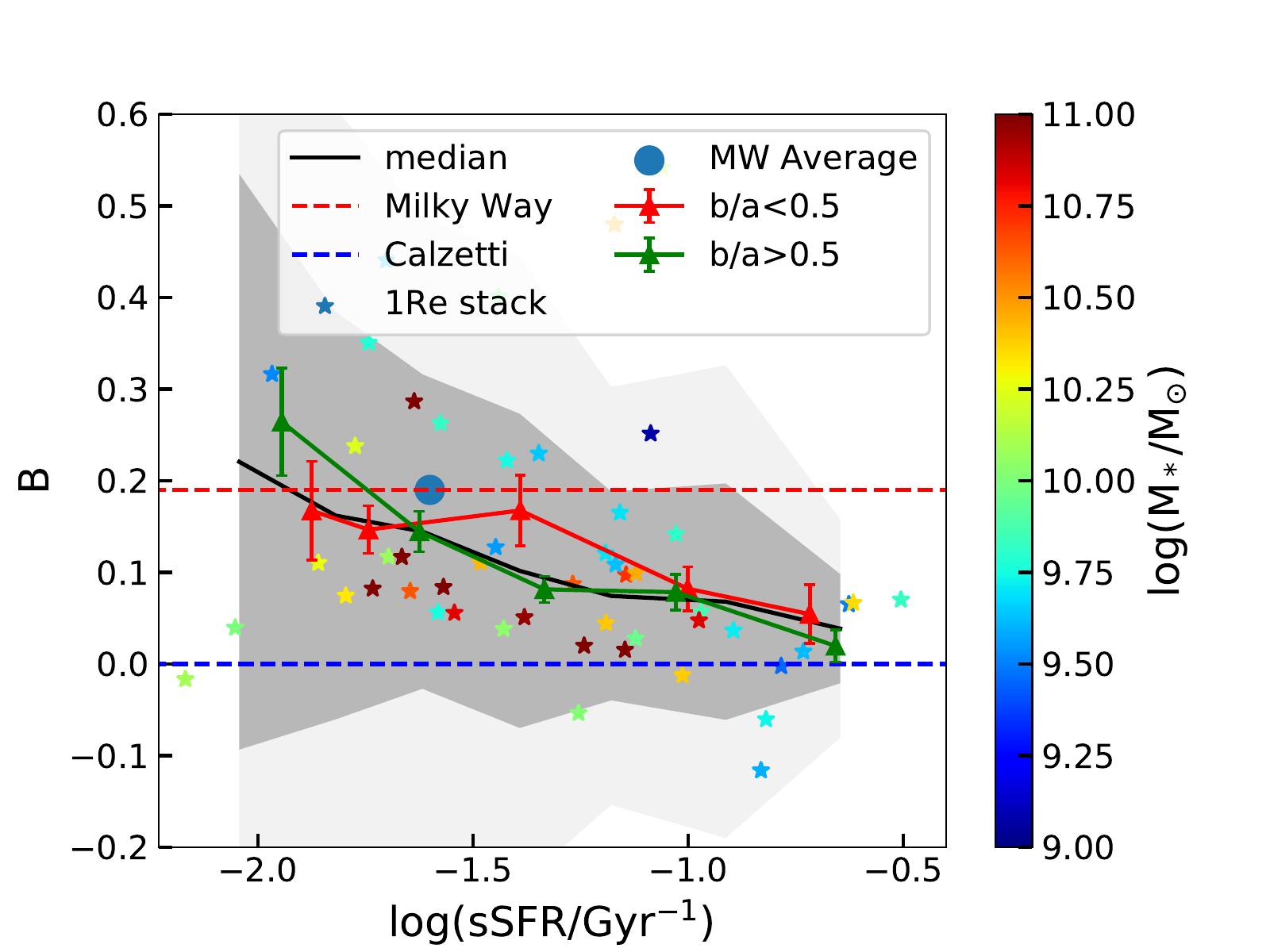}
	\includegraphics[width=0.33\textwidth,height=0.25\textwidth]{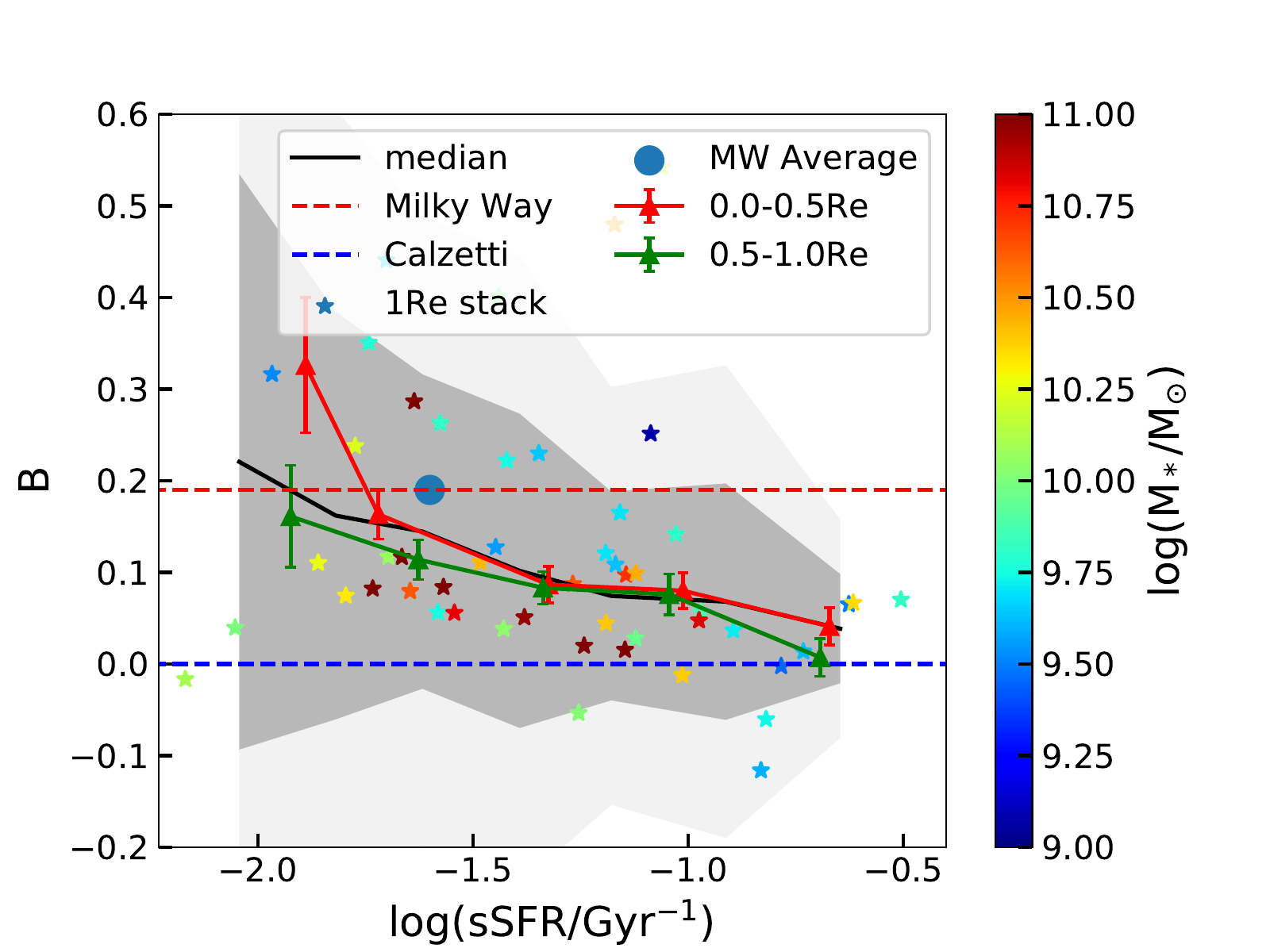}
 
	\caption{The attenuation excess due to the 2175{\AA} bump ($A_{\rm bump}$, top) and normalized bump strength ($B$, bottom)  in our sample as a function of sSFR. The black line shows the median values of the results obtained from individual pixels, with grey shaded regions indicates 1$\sigma$ and 2$\sigma$ scatters of the data points.
		Stars are results from the 1Re stacks of each galaxy, with color codes showing the total stellar mass of the galaxy (from NSA).
		 In panels from left to right, colored triangles/lines show results of subsamples of spaxels selected by \Av\ (left), $b/a$ (middle) and the galactocentric distance (right), respectively. In bottom panels,
		the bump strength for the Milky-Way and Calzetti dust curve are indicated as red and blue dash lines respectively, with a blue circle indicating the average value of the Milky Way ($\log(\rm sSFR/Gyr^{-1})\sim -1.6$).}
	\label{fig:sSFRvsB}
\end{figure*}

\begin{figure*}
	\includegraphics[width=0.33\textwidth]{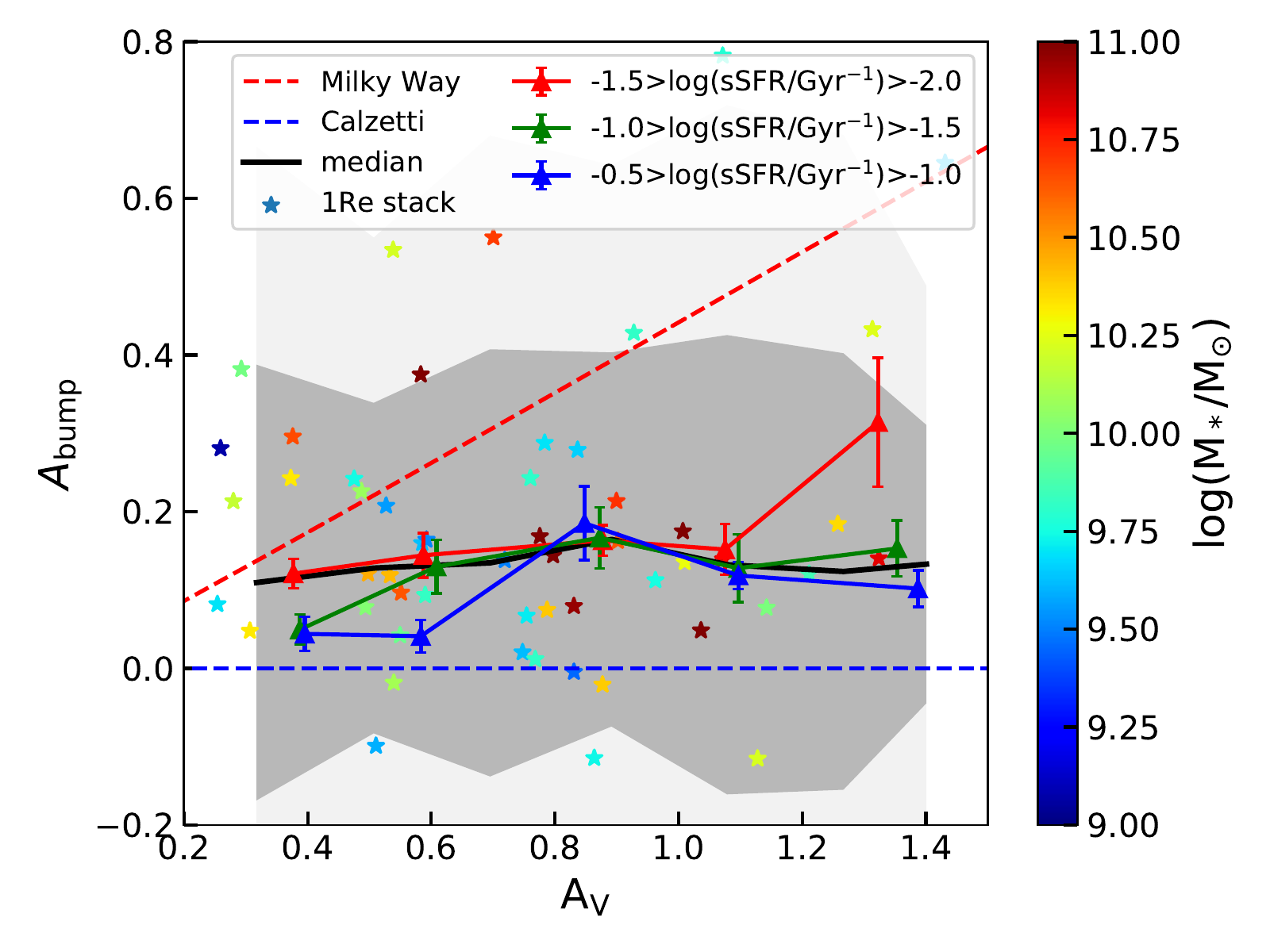}
	\includegraphics[width=0.33\textwidth]{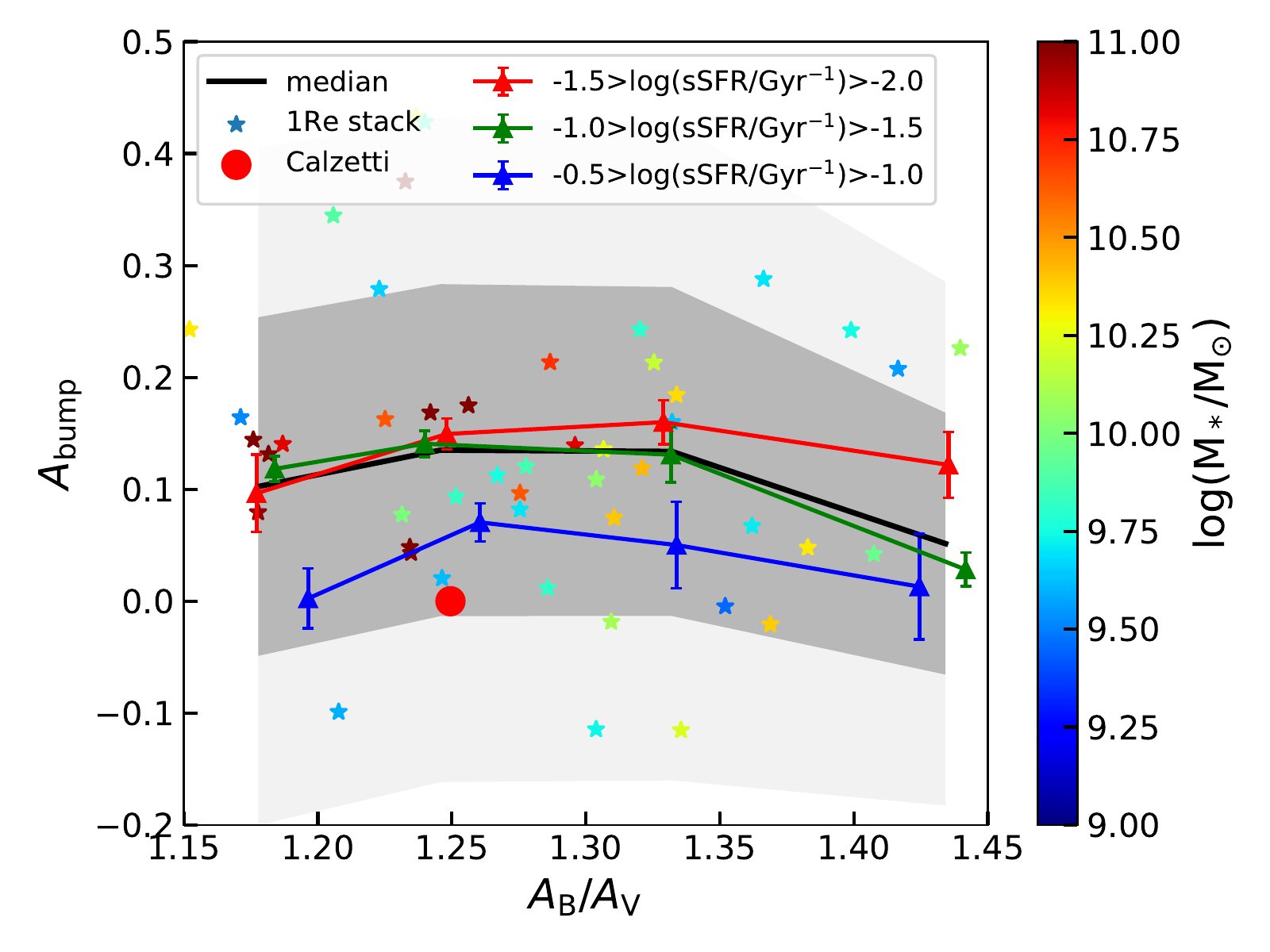}
	\includegraphics[width=0.33\textwidth]{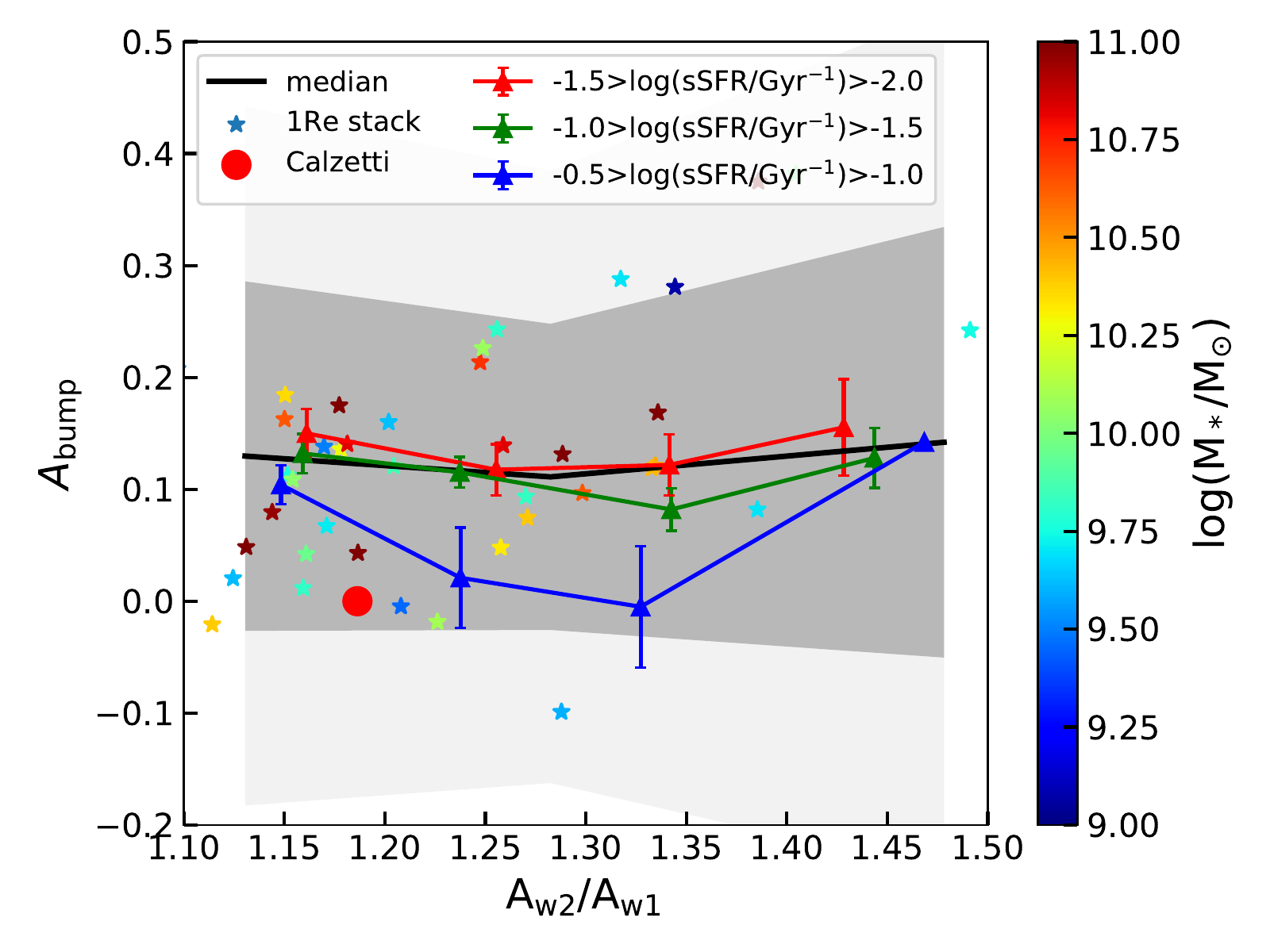}\\  
 	\includegraphics[width=0.33\textwidth]{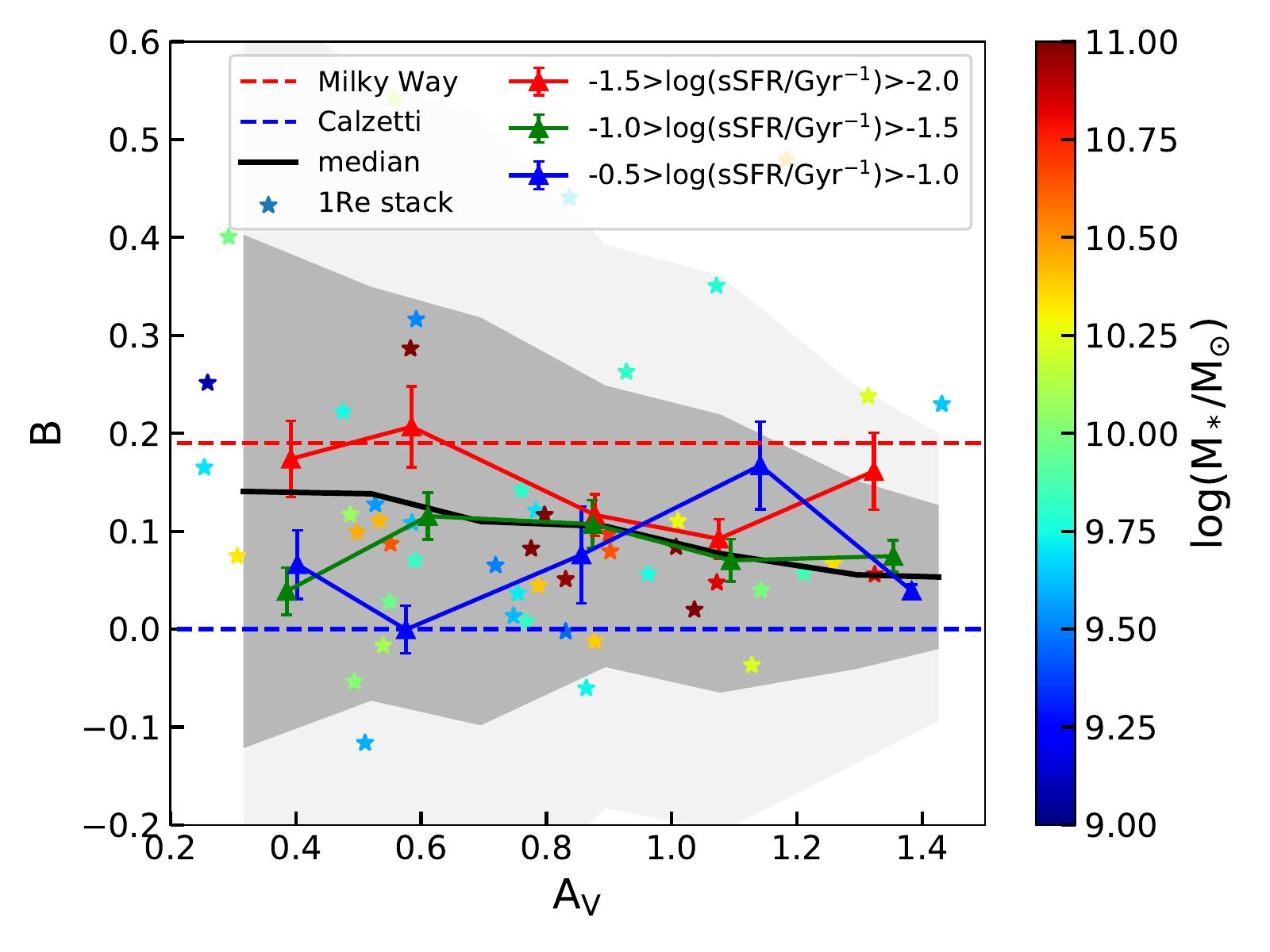}
	\includegraphics[width=0.33\textwidth]{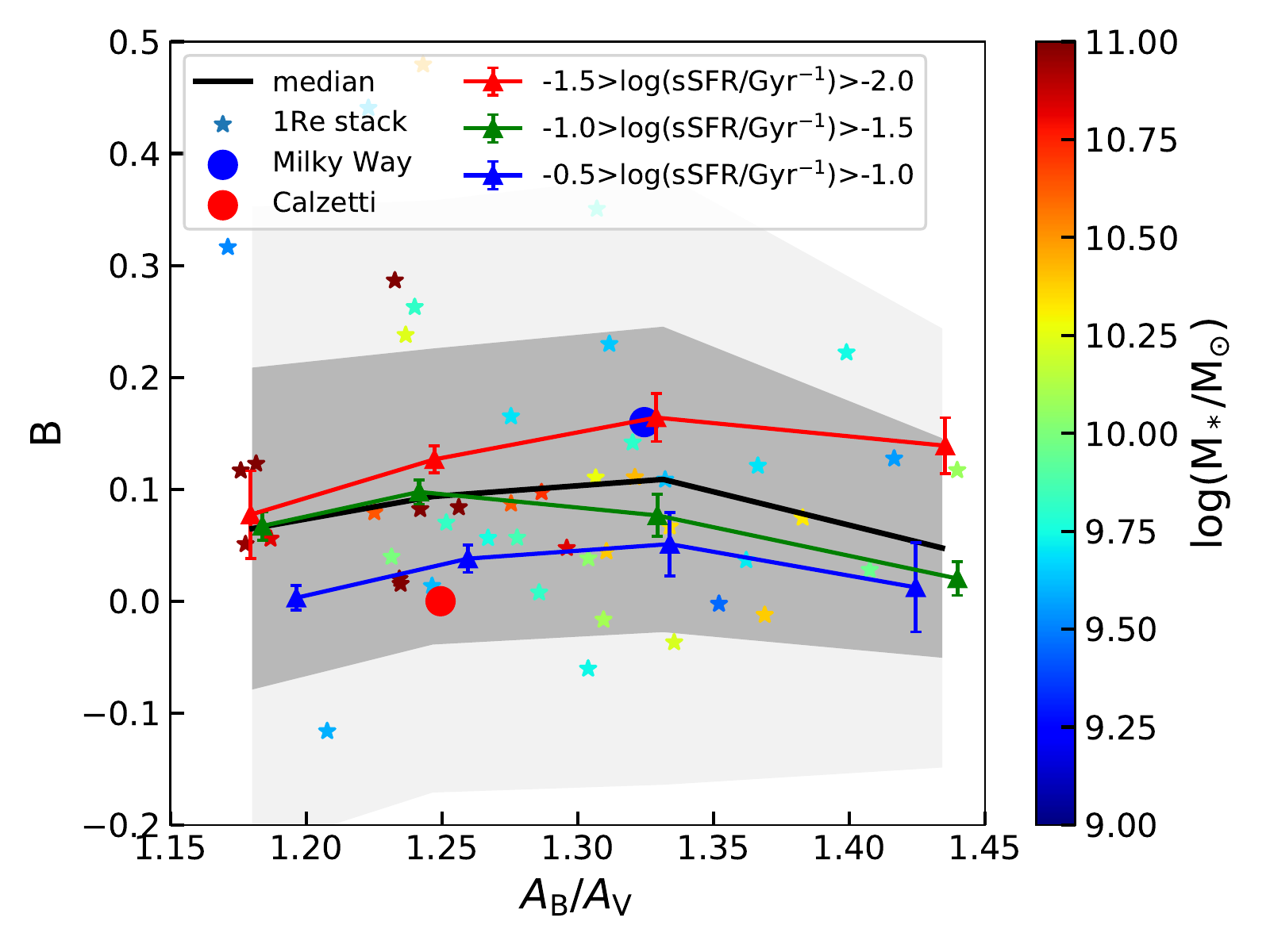}
	\includegraphics[width=0.33\textwidth]{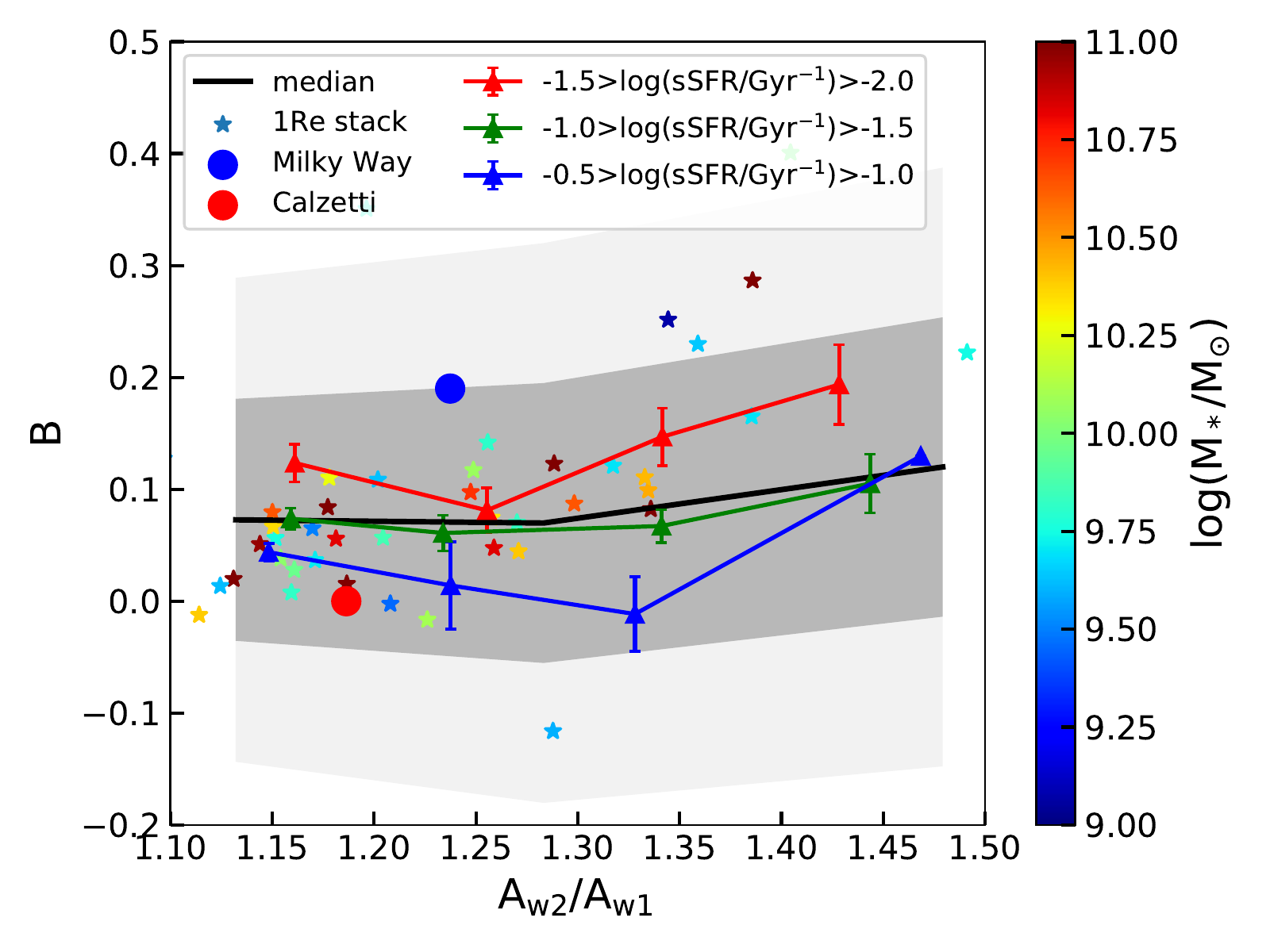}
	\caption{The attenuation caused by the 2175{\AA} bump ($A_{\rm bump}$, top) and normalized bump strength ($B$, bottom)  in our sample as a function of \Av (left) and the slopes of the attenuation curve in optical ($A_{\rm B}/A_{\rm V}$, middle) and NUV ($A_{\tt w2}/A_{\tt w1}$, right).
		The black line shows the median values of the results obtained from individual pixels, with grey shaded regions indicating 1$\sigma$ and 2$\sigma$ scatters of the data points. Stars are results from the 1Re stacks of each galaxy, with color codes showing the total stellar mass of the galaxy (from NSA). In panels from left to right, colored triangles/lines show results of subsamples of spaxels selected by \Av\ (left), $b/a$ (middle) and the galactocentric distance (right), respectively. Values for the Milky-Way and Calzetti dust curve are indicated in red and blue respectively.}
	\label{fig:Bvslope}
\end{figure*}

To see the \Av\ dependence of the slope more clearly,  
in \autoref{fig:curve_relation_AV} we plot $A_{\rm B}/A_{\rm V}$ (left panel) 
and $A_{\tt  w2}/A_{\tt w1}$ (middle panel) as functions of \Av, 
as well as the relation between $A_{\rm B}/A_{\rm V}$ and $A_{\tt  w2}/A_{\tt w1}$
themselves (right panel). Results for three subsamples selected by 
$\log_{10}{\rm sSFR}$ are plotted as colored triangles connected by lines.
On average the slopes in both optical and NUV are anti-correlated with \Av.  
In both the optical and NUV bands, a Calzetti slope is preferred by spaxels/galaxies
with \Av$\sim0.9$ mag.  The slope becomes comparable to or steeper than 
the Milky-Way type at lower \Av, and flatter than the Calzetti curve at higher \Av. 
At \Av$\ga0.8$, the slope-\Av\ relation remain similar even when spaxels 
are divided into narrow ranges of sSFR, suggesting that the slope is 
primarily determined by \Av\ rather than by sSFR in regions of high dust opacity. 
At \Av$\lesssim0.8$, the anti-correlations between the slope 
and \Av\ may be partly attributed to the positive correlation 
between \Av\ and sSFR at the low ends of opacity and sSFR (see \autoref{fig:AV_sSFR}). Also note that at the low ends of \Av\ the scatters of the correlation are significantly larger than at the higher end, especially in the left panel. This is expected as the uncertainty of $A_{\rm B}/A_{\rm V}$ directly correlates with the absolute value of \Av\, and the results again signify the importance of excluding spaxels of very low  \Av\ in our analysis.
Our results confirm that a Calzetti curve is valid for star-forming galaxies 
with large dust contents, while dust curves of steeper slopes may be 
needed for galaxies with low optical opacity. 

 The right panel of \autoref{fig:curve_relation_AV} shows 
that the two slopes are positively and (almost) linearly correlated with 
each other, with scatter that is comparable 
to that seen in the test (\S~\ref{sec:tests}). Both the Milky Way and
Calzetti curves lie on the median relation of our sample, but 
the variance among individual galaxies is large. 
This again suggests that dust attenuation laws in galaxies cannot be fully 
characterized by curves with a single slope. These results are consistent with 
those obtained previously \citep[e.g.][]{Salim2018}, and with  
radiative transfer model predictions taking into account complex  
dust-gas distributions \citep{Narayanan2018}. Since previous observations 
have been mostly limited to global relations of galaxies, our results 
show for the first time that these correlations also hold 
for regions of kpc scales within individual galaxies. 
As can be seen, the median relations remain almost unchanged 
when the sSFR is limited to narrow ranges, consistent with the weak 
dependence of the slope on sSFR seen in \autoref{fig:slope_opt_sSFR} 
and  \autoref{fig:slope_NUV_sSFR}.

\subsection{The 2175{\AA} bump}

In \autoref{fig:sSFRvsB}, we plot the attenuation excess due 
	to the 2175{\AA} bump ($A_{\rm bump}$, top) and normalized bump strength 
	($B$, bottom), estimated for spaxels and galaxies as a function of $\log_{10}(\rm sSFR)$.
The results for spaxels as a whole and for individual galaxies
are plotted with the same symbols/lines coding as in the previous figures,
and are repeated in the three panels that show results for subsamples 
selected by \Av \ (left), $b/a$ (middle) and galactocentric distance 
(right). The values of $B$ obtained from our samples range from 
$\sim0$ to $\sim0.5$, broadly consistent with the strengths reported 
for galaxies at high redshift ($1<z<2$, \citealt{Buat2011}; $0.5<z<2$, \citealt{Kriek2013}; $z\sim2$, \citealt{Shivaei2022}). 
Both $A_{\rm bump}$ and $B$ decrease with increasing sSFR, with a median bump strength 
that is comparable to or stronger than that of the Milky Way-type bump
in quiescent regions with $\log_{10}({\rm sSFR}/{\rm Gyr^{-1}})\lesssim-1.8$, and 
consistent with the Calzetti curve with $B=0$ in star-forming regions 
with the highest sSFRs.  This result is also consistent with previous findings 
based on global measurements of high-$z$ galaxies 
\citep[e.g.][]{Kriek2013,Kashino2021}, as well as with the absence of 
the 2175 {\AA} bump in local starburst galaxies used to
derive the Calzetti curve \citep{Calzetti1994,Calzetti2000}.

From \autoref{fig:sSFRvsB} and for both the $A_{\rm bump}$-sSFR 
	relation and the $B$-sSFR relation, we find no obvious dependence 
	on $A_V$, $b/a$ and the galactocentric distance in all the panels
	except the top-left panel where the spaxels with $A_V<0.6$ and 
	$\log_{10}({\rm sSFR/Gyr^{-1}})\ga -1.7$ appear to have smaller 
	$A_{\rm bump}$ at fixed sSFR than those with higher optical opacities. 
	This result indicates that 
the properties responsible for the variation of the 2175{\AA} bump 
in galaxies are not sensitive to the total attenuation, global 
inclination of galaxies and the location within individual galaxies. 
Instead, the variation may be related to processes on scales 
smaller than the sizes of individual spaxels (kpc scales) studied here.

In \autoref{fig:Bvslope}, we further examine $A_{\rm bump}$ 
	(upper panels) and $B$ (lower panels) as a function of \Av\ (left panels), 
	$A_{\rm B}/A_{\rm V}$ (middle panels), and $A_{\rm w2}/A_{\rm w1}$ (right panels). 
	As in previous figures, in each panel the black line and the shaded 
		regions present the median relation and the $1\sigma$ and $2\sigma$ 
		scatter of individual spaxels around the median, while 
		the colored triangles connected by the solid lines display 
		the results for subsamples selected by sSFR.		
		The red and blue dashed lines in the left panels
		indicate the relations for the Milky Way-like and Calzetti attenuation 
		curves. If the attenuation curve has a fixed Milky Way-like shape, 
		one would expect a fixed bump strength ($B=0.19$) and 
		a linearly increasing $A_{\rm bump}$ with increasing \Av.
		This correlation is not seen in our result, however. As can be seen
		from the upper-left panel, $A_{\rm bump}$ at scales of both 
		spaxels and galaxies shows no siginificant correlation with \Av\
	    and this result is held at limited ranges of sSFR.
	In the lower-left panel the median $B$ parameter of all the 
	spaxels tends to present a marginal anti-correlation with \Av, 	
	which becomes rather insigificant when the spaxels are divided 
	into sSFR intervals. In the middle and right panels, 
	$A_{\rm bump}$ and $B$ show no  obvious correlation with the 
	attenuation curve slope in both optical and NUV bands, which is 
	similar to the results reported by \cite{Barisic-2020} based on 
	spectroscopy of galaxies at $z\sim0.8$ from the LEGA-C survey.

	\autoref{fig:Bvslope} also shows that the overall scatter of the bump 
	properties at fixed  $A_{\rm B}/A_{\rm V}$ or  $A_{\rm w2}/A_{\rm w1}$
	 can be largely explained by the correlation of the bump parameters 
	 with sSFR as seen in \autoref{fig:sSFRvsB}. 
    Combined with the null correlation of the bump properties with 
    the attenuation curve parameters, this result indicates that the variation 
    of the 2175\AA\ bump and the variation of the attenuation curve are 
    unlikely to be driven by the same process. It is likely that 
    the variation of the 2175{\AA} bump strength is driven by star 
    formation-related processes, such as the destruction of the bump 
    carriers by the UV radiation produced by newly formed 
    massive stars \citep[e.g.][]{Fischera2011}. We will come back and 
    discuss this point in the next section.


\section{Discussion}
\label{sec:discussion}

\subsection{Methods of deriving dust attenuation curves}

In order to derive the attenuation curve for a galaxy or a resolved region, 
one has to obtain the intrinsic, dust-free spectrum or SED in the first place. 
In previous studies this was done usually in two different ways, either by 
empirically comparing observed SEDs/spectra between dusty galaxies and 
those with no or relatively weak attenuation
\citep[e.g.][]{Calzetti1994,Kinney-1994,Calzetti1997,Calzetti2000,
	Johnson-2007,Wild2011,Battisti2016,Battisti2017}, 
or based on stellar population synthesis (SPS) models 
\citep[e.g.][]{Spinrad-1971,Faber-1972,Sawicki-Yee-1998,Papovich-2001,
Kauffmann2003,Salim-2005,Salim-2007,Noll2009,Conroy2013,Leja2017,
Boquien2019,Johnson-2019,Zhou2019,Boquien-2022,Jones-2022,Nagaraj-2022}.

As pointed out by \cite{Wild2011}, in the empirical method the template or 
the galaxy in a pair with higher attenuation will contribute more to the final 
average curve if the curve 
slope depends on the dust attenuation itself, as is observed in previous studies 
and in \autoref{fig:curve_relation_AV}. Since the dust content varies across a 
galaxy and the dust curve slopes span a wide range, as revealed by 
our spatially resolved measurements (see \autoref{fig:example} for examples), 
it is no longer appropriate to assume a 
single form of attenuation to rank templates or paired galaxies
\citep{Salim2020}.

The SED-fitting method allows attenuation curves 
to be obtained in a parametric form, but this method suffers from the well-known 
dust-age-metallicity degeneracy which may be alleviated by using  
IR data \citep{Burgarella-2005}. More recently it is found that dust attenuation curves can 
be constrained even without IR data \citep[e.g.][]{Kriek2013,Salim-2016,Salim2018}.
However, by examining systematic biases in dust attenuation curves derived 
from SED modelling, \cite{Qin-2022} concluded that the relation between the slope of dust 
attenuation curve and the optical depth found previously 
is likely caused by the degeneracy between model parameters.
\cite{Lower-2022} proposed a new method using a nonuniform 
screen dust model to allow a fraction of a galaxy to be unattenuated,
and found a significant improvement in the dust attenuation modelling accuracy.
Their finding again demonstrated the importance of using spatially resolved data. 

Our method belongs to the second category.
In our method we obtain the attenuation curve in the optical before applying
our spectral fitting code {\tt BIGS} to determine the stellar population
properties, thus largely reducing the influence of the degeneracy between 
model parameters. The relation between the slope of dust attenuation curve and 
the optical depth is clearly seen in our sample (see \autoref{fig:curve_relation_AV}), 
suggesting that this relation is not entirely a result of systematic bias caused 
by degeneracy of model parameters as found in \citet{Qin-2022}. In addition, by applying our 
method to spatially resolved data, we have naturally taken into account 
the non-uniform distribution of dust attenuation in galaxies, thus achieving 
higher accuracy in the dust attenuation measurements as found in 
\citet{Lower-2022}. Furthermore, by using NIR photometry to calibrate the 
absolute amplitude of the attenuation curves, we have taken a step forward 
in using the technique of \citet{Li2020} which provides only relative attenuation curves.
Finally, by combining optical spectra from MaNGA and NUV photometry
from {\it Swift}/UVOT, we have developed a novel method to estimate the 
2175\AA\ bump in attenuation curves. Clearly, our method can be tested and 
improved further using larger samples with broader wavelength coverage.

\subsection{Comparison with previous observations}

In the local Universe, 
previous studies have revealed a wide range of the attenuation curve slope, from those 
similar to the Calzetti curve to those steeper than the Milky Way 
extinction curve,  as well as a trend of the curve slope with 
the optical opacity in the sense that galaxies with higher $A_V$ tend 
to have shallower slopes.
\citep[e.g.][]{Burgarella-2005,Conroy2010dust,Wild2011,Battisti2016,Battisti2017,Leja2017,Salim2018,Boquien-2022,Nagaraj-2022}.
In particular, using a sample of 23,000 galaxies
with data from {\it GALEX}, SDSS and WISE, \citet{Salim2020} showed 
that, once $A_V$ is fixed, the slope-$A_V$ relation does not show any 
significant residual dependence on the stellar mass, axis ratio, sSFR and 
stellar metallicity of the galaxies. The lack of dependence on stellar 
mass and sSFR was noticed by some earlier studies based on empirical 
methods \citep[e.g.][]{Johnson-2007,Wild2011,Seon2016}.
In our work, the slope-$A_V$ anti-correlation and the lack of dependence 
on stellar mass, axis ratio and sSFR found earlier on galactic scales
are also clearly seen at kpc scales 
(Figs.~\ref{fig:slope_opt_sSFR},~\ref{fig:slope_NUV_sSFR} and~\ref{fig:curve_relation_AV}),
indicating that the global trends found previously originate 
from local regions at scales of kpc or smaller.

For the UV bump at 2175\AA, previous studies of local galaxies based on 
the empirical method have not led to clear conclusions: with no/weak bump in some studies 
\citep[e.g.][]{Kinney-1993,Calzetti1994,Calzetti2000,Battisti2016,Battisti2017}
and a moderate bump in others \citep[e.g.][]{Wild2011}.
In contrast, most of the studies based on SED-fitting techniques 
found evidence for a bump, with the strength ranging 
from zero to values close to or even higher than the MW's bump 
\citep[e.g.][]{Burgarella-2005,Conroy2010dust,Salim2018}. The 
results from \citet{Salim2018} supported the existence of a correlation
between the curve slope and the bump strength, with weaker bumps
in shallower curves. A similar trend has also been found for high-z galaxies
\citep[e.g.][]{Kriek2013}. However, this correlation is not clearly 
seen in our result. As can be seen in Fig.~\ref{fig:Bvslope}, the bump strength 
shows no correlation with $A_{\rm B}/A_{\rm V}$; it seems to 
slightly increase with the NUV slope, but the trend is rather week.
Note that in \cite{Kriek2013}, the bump strength is tied to the slope of the curve by definition, which is different from other works in which a bump profile is simply added to the bump-free attenuation curve \citep[e.g.][]{Barisic-2020,Kashino2021,Shivaei2022}. Such a definition may induce a correlation between the bump strength and the curve slopes. Our method here does not rely on any specific attenuation curve, with the bump strength calculated in the $uvm2$ band, which can also introduce minor differences in the derived properties. Cautions are urged  when comparing different bump strength measurements. Apart from the detail differences, the 
 strongest correlation we have found for the bump strength is 
with the sSFR (see Fig.~\ref{fig:sSFRvsB}), with $B$ decreasing with 
increasing sSFR, and this correlation holds even when other properties 
are limited to narrow ranges. The same trend 
was previously reported by \citet{Kriek2013} and \citet{Kashino2021} 
for high-z galaxies based on global SED fitting.
\citet{Shivaei2022} reported a similar but more complex trend such that the UV bump amplitude increases with mass at fixed SFR but does not change with SFR at fixed stellar mass.
Again, our results indicate that the global trend seen previously is a result of 
the local trend present on kpc or smaller scales. 

Our study is the first attempt to derive spatially-resolved dust attenuation 
curves and the UV bump strength for a sample of galaxies beyond the 
Local Group, though still with limited sample size (72 galaxies) and 
spatial resolution ($\sim$kpc). We note that the {\it Swift}/UVOT data have 
been used to study the spatial resolved dust attenuation curve and 
UV bump for a handful of nearby galaxies, e.g. 
M81 and Holmberg IX by \citet{Hoversten-2011}, SMC by \citet{Hagen2017} and NGC 628 by \citet{Decleir2019}. Both M81 and Holmberg IX can be  best fit with 
a Milky Way extinction curve with a prominent UV bump,
while the median attenuation curve of NGC 628 and SMC is fairly steep, with 
a sub-MW-type UV bump. In addition, intriguing variations are observed 
among different regions in NGC 628 and SMC, with regions of higher 
$A_{\rm V}$ found to have shallower attenuation curves and weaker UV bumps.
Taking advantage of the 
wide ranges of galaxy properties covered by our sample, our study confirms those findings in case studies and investigate further the $B$-sSFR relation which was not   seen clearly due to the lack of spatially
resolved spectroscopy and of a large sample.

\subsection{What drives the variation in the 2175{\AA} bump?}

Although the exact carriers of the UV bump are still under debate, most of the 
studies agree that they should be some kind of carbonaceous grains with small 
sizes ($\lesssim 0.01\mu m$, \citealt{Fischera2011}). The production and destruction 
of the dust grains can thus affect the observed UV bump strength.
In fact, many investigations went along this line to study the 
grain size distribution, and try to correlate the distribution with observed
attenuation curves \citep[e.g.][]{Fischera2011,Asano2013,Hirashita2020}. 
For instance, \cite{Fischera2011} proposed that the ambient UV radiation field 
can destroy the carriers of the UV bump, and found that their models can naturally explain 
both the shallow slope and the absence of the UV bump for the Calzetti curve. 
\cite{Hirashita2020} modelled the grain population by considering
various dust components: silicate, aromatic carbon, and non-aromatic carbon, 
and found that their model can reproduce both the Milky Way and Calzetti curves by 
varying the star formation time scale and the dense gas fraction.
Our finding that regions with higher sSFRs tend to have weaker UV bumps 
supports this scenario, as regions of intensive star formation produce strong UV 
photons that may destroy the bump carriers and thus reduce the 
bump strength. In the recent work of \citet{Kashino2021}, where 
an anti-correlation between the bump strength and sSFR was also 
found for galaxies at $z\sim0.8$, the bump variation was suggested to be
determined by the recent star formation history of galaxies through the 
destruction of small carbonaceous grains by supernovae and intense 
radiation fields. A more recent work by \cite{Shivaei2022} found that the  UV bump strength   highly correlates with the PAH strength indicated by the Spitzer MIPS 24 $\mu m$ photometry, and that younger galaxies or galaxies that have experienced a recent (10-100 Myrs ago) starburst would have both reduced bump strengths and elevated mid-IR emission of PAHs.

On the other hand, even in the absence of the variation in the underlying dust 
grain properties and in the intrinsic dust extinction curve, 
the observed attenuation curve may still vary due to radiative transfer effects.
For instance, 
\cite{Seon2016} modelled the dust attenuation curves in a clumpy interstellar medium,
and found that the attenuation curves are not necessarily determined by the underlying 
extinction curve. Instead, the absorption or scattering efficiency affects the shape 
of the curve, and the 2175{\AA} bump from a MW extinction curve can be suppressed 
by using different albedos. Similarly, \cite{Narayanan2018} combined the 
dust radiative transfer model with cosmological zoom-in simulations to show that 
the star-to-dust geometry alone can lead to the absence of the bump in the attenuation 
curve of a galaxy. Moreover, although different models give different 
explanations for the varying bump strength, they all predict flatter 
attenuation curves and weaker UV bumps for higher optical opacity. 
Since this prediction was in line with their findings 
for high-z galaxies at $0.5<z<2$, \cite{Kriek2013} suggested that the observed 
trends can be explained by differences in star-dust geometry, a varying 
grain size distribution, or both. 
By comparing galxies in bins of sSFR and $A_{\rm V}$ in our sample, we found that the UV bump strength is more fundamentally anti-correlate with sSFR, while no obviously correlations with $A_{\rm V}$ are seen. In this regard, 
our result is actually not in favor of star-dust geometry as the driver for 
the UV bump variation. In addition, by using spatially-resolved data we have 
largely reduced the effect of the star-dust geometry, although the spatial resolution 
of our data is not high enough to fully resolve star-forming regions. 
Therefore, we argue that the variation of the UV bump can be more naturally 
explained by the first scenario, that is, destruction of dust grains to varying 
degrees due to varying UV photons produced by different star formation activities.
Data with higher resolutions are needed to discriminate between
different scenarios at sub-kpc scales, given that radiative transfer 
and geometry effects cannot be ignored at the MaNGA resolution.

\section{Summary}
\label{sec:summary}

We have developed a novel method to measure the optical and NUV slopes 
as well as the 2175{\AA} bump of dust attenuation curves at kpc scales for a sample of 72 galaxies 
in the local Universe, which have integral field spectroscopy from the  
MaNGA survey, NIR photometry in Ks band from 2MASS, and {\it Swift}/UVOT
NUV photometry in {\tt uvw2}, {\tt uvm2} and {\tt uvw1} bands from the 
SwiM catalog. For a given region in our sample, we first apply the technique 
of \cite{Li2020} to the MaNGA spectrum to derive a model-independent 
relative attenuation curve in the optical range ($A_\lambda-A_{\rm V}$). 
Next, the observed spectrum from MaNGA is corrected for dust attenuation 
using the calibrated attenuation curve, and is fitted with our Bayesian 
inference code {\tt BIGS} for stellar population synthesis to derive a dust-free 
model spectrum which extends to NUV and NIR. Finally, the dust-free spectrum is 
compared with the observed optical spectrum, 2MASS and the {\it Swift}/UVOT photometry, 
yielding the total attenuation in different bands
($A_{\rm B}$, \Av, $A_{\rm w2}$, $A_{\rm m2}$, $A_{\rm w1}$),
the slope of the attenuation curve in both optical ($A_{\rm B}/A_{\rm V}$) 
and NUV ($A_{\rm w2}/A_{\rm w1}$), and the strength of the 2175{\AA} bump 
($B$). We have applied this method to a set of mock spectra/SEDs 
with different SNRs, which are generated to cover a wide parameter space
in stellar population and dust attenuation properties and mimic the quality
of the data used in our study. This test shows that our method is able to well 
reproduce the average properties of dust attenuation in individual regions of 
our galaxies, with no/weak systematics and reasonably small errors.

We have examined the correlations of the dust attenuation parameters as 
measured in individual regions with the specific star formation rate (sSFR),
as well as the correlations between the attenuation parameters themselves. 
We have also obtained {\em global} measurements of the dust attenuation
parameters of individual galaxies, by applying the same method to the 
stacked spectrum and total photometry within the effective radius ($R_e$)
of each galaxy. These measurements allow us to examine possible 
dependence of the dust attenuation laws on galaxy mass.

Our main results can be summarized as follows.
\begin{itemize}
	\item  Attenuation curves at kpc scales span a wide range of slopes in 
	both optical and NUV,  from those shallower than the Calzetti curve to 
	those steeper than the Milky Way-type curves.
	\item The slope in both optical and NUV becomes shallower as one goes 
	from low to high optical opacities, and this trend remains almost unchanged 
	when the sample is limited to narrow ranges of sSFR, $b/a$, or the location 
	within host galaxies.
	\item The 2175\AA\ UV bump in regions of kpc scales  presents a wide range 
	of strengths, ranging from that of a Calzetti curve with no bump 
	to those stronger than the bump in Milky Way-type curves.
	\item The 2175\AA\ bump strength is primarily driven by sSFR with weaker 
	bumps at higher sSFR, but not by the optical opacity, $b/a$ and the location
	within host galaxies.
	\item The correlation between the  2175\AA\ bump strength with sSFR instead of optical opacities  strongly suggests that the 
	UV bump variation is more fundamentally driven by processes related 
	to star formation 
	(e.g. destruction of dust grains by UV radiation in strongly star-forming regions).
	\item The above trends appear to be independent of the stellar mass of 
	galaxies, indicating that the dust attenuation laws are driven by local 
	processes rather than by global properties of galaxies.
\end{itemize}

\section*{Acknowledgments}

This work is supported by the National Key R\&D Program of China
(grant No. 2018YFA0404502), and the National Science
Foundation of China (grant Nos. 11821303, 11733002, 11973030,
11673015, 11733004, 11761131004, 11761141012). ME and CG acknowledge support from NASA grant 80NSSC20K0436. The work of MM is supported in part through a fellowship sponsored by the Willard L. Eccles Foundation. RY acknowledges support by the Hong Kong Global STEM Scholar Scheme, by the Hong Kong Jockey Club through the JC STEM Lab of Astronomical Instrumentation program, and by a grant from the Research Grant Council of the Hong Kong (Project No. 14302522).

Funding for SDSS-IV has been provided by the Alfred P. Sloan Foundation and
Participating Institutions. Additional funding towards SDSS-IV has been provided
by the US Department of Energy Office of Science. SDSS-IV acknowledges support
and resources from the Centre for High-Performance Computing at the University
of Utah. The SDSS web site is www.sdss.org.

SDSS-IV is managed by the Astrophysical Research Consortium for the Participating
Institutions of the SDSS Collaboration including the Brazilian Participation Group,
the Carnegie Institution for Science, Carnegie Mellon University, the Chilean
Participation Group, the French Participation Group, Harvard–Smithsonian Center
for Astrophysics, Instituto de Astrofsica de Canarias, The Johns Hopkins University,
Kavli Institute for the Physics and Mathematics of the Universe (IPMU)/University
of Tokyo, Lawrence Berkeley National Laboratory, Leibniz Institut fur Astrophysik
Potsdam (AIP), Max-Planck-Institut fur Astronomie (MPIA Hei- delberg),
Max-Planck-Institut fur Astrophysik (MPA Garching), Max-Planck-Institut fur
Extraterrestrische Physik (MPE), National Astronomical Observatory of China,
New Mexico State University, New York University, University of Notre Dame,
Observatario Nacional/MCTI, The Ohio State University, Pennsylvania State University,
Shanghai Astronomical Observatory, United Kingdom Participation Group, Universidad
Nacional Autonoma de Mexico, University of Arizona, University of Colorado Boulder,
University of Oxford, University of Portsmouth, University of Utah, University of
Virginia, University of Washington, University of Wisconsin, Vanderbilt University
and Yale University.

\bibliographystyle{aasjournal}

\bibliography{szhou}

\end{document}